\documentclass[12pt]{article}
\usepackage{amsmath,amssymb}
\usepackage{graphicx,psfrag,epsf}
\usepackage{enumerate}
\usepackage{natbib}
\usepackage{caption}
\usepackage{subcaption}
\usepackage[colorlinks, citecolor = blue, linkcolor = blue]{hyperref}
\usepackage{centernot}
\usepackage{authblk}
\usepackage{url}  
\usepackage{bm}
\usepackage{floatrow}
\usepackage{microtype}
\usepackage{enumitem}
\usepackage{mathrsfs}

\usepackage{bigints}
\usepackage{float}
\usepackage{amsthm}
\usepackage[ruled,vlined]{algorithm2e}
\usepackage[english]{babel}

\newcommand{\tR}{\mathbb{R}}

\def\E{\mathbb{E}}

\newcommand{\blind}{1}
\newfloatcommand{capbtabbox}{table}[][\FBwidth]

\usepackage{tikz}
\usetikzlibrary{shapes,arrows}
\usepackage{tikzsymbols}

\begin{document}
\tikzstyle{block} = [rectangle, draw, fill=blue!20, 
text width=5em, text centered, rounded corners, minimum height=4em]
\tikzstyle{line} = [draw, -latex']
\tikzstyle{cloud} = [draw, ellipse,fill=red!20, node distance=3cm,
minimum height=2em]

\def\spacingset#1{\renewcommand{\baselinestretch}%
	{#1}\small\normalsize} \spacingset{1}

\if1\blind
{
	\title{\bf Forward Stability and Model Path Selection}
	\author[1]{Nicholas Kissel}
	\author[2]{Lucas Mentch}
	\affil[1]{\small Department of Statistics and Data Science, Carnegie Mellon University}
	\affil[2]{\small Department of Statistics, University of Pittsburgh}
	\maketitle
} \fi

\if0\blind
{
	\bigskip
	\bigskip
	\bigskip
	\begin{center}
		{\LARGE\bf Title}
	\end{center}
	\medskip
} \fi

\begin{abstract}
\noindent Most scientific publications follow the familiar recipe of (i) obtain data, (ii) fit a model, and (iii) comment on the scientific relevance of the effects of particular covariates in that model. This approach, however, ignores the fact that there may exist a multitude of similarly-accurate models in which the implied effects of individual covariates may be vastly different. This problem of finding an entire collection of plausible models has also received relatively little attention in the statistics community, with nearly all of the proposed methodologies being narrowly tailored to a particular model class and/or requiring an exhaustive search over all possible models, making them largely infeasible in the current big data era. This work develops the idea of forward stability and proposes a novel, computationally-efficient approach to finding collections of accurate models we refer to as model path selection (MPS). MPS builds up a plausible model collection via a forward selection approach and is entirely agnostic to the model class and loss function employed. The resulting model collection can be displayed in a simple and intuitive graphical fashion, easily allowing practitioners to visualize whether some covariates can be swapped for others with minimal loss.
\end{abstract}

\noindent%
{\it Keywords:} Stability Selection, Ranking and Selection, Rashomon Effect

\spacingset{1.45}

\section{Introduction}
\label{sec:intro}

Despite the rapid acceleration into the modern data science era, many practitioners remain tethered to a classical approach to data analyses.  In regression problems, such an approach can be broadly characterized by the following steps:  (i) obtain data, (ii) choose a model class, often based largely on tradition or personal preference and with no or only minimal heuristic justification, (iii) use the data and a selection method to obtain the empirically optimal model within that class, and (iv) devote the remaining discussion to the broader scientific implications of the particular model selected, paying particular attention to the specific covariates that are selected or omitted and their implied relationship to the response.

More formally, given data $Z_1, ..., Z_n \sim F_Z$ in the form of ordered pairs $Z=(X,Y)$ consisting of covariates $X = (X_1, ..., X_p)$ and a response $Y$, we commonly imagine a generic regression relationship of the form $Y = f(X) + \epsilon$.  To perform the regression, one often begins with a collection of models $\mathcal{M} = \{f_{\lambda}: \lambda \in \Lambda \}$ indexed by some parameter $\lambda \in \Lambda$ that often serves to enforce some kind of regularization such as the number of steps in forward selection or the amount of shrinkage in penalized techniques like the lasso. Model selection procedures are designed to select some $\lambda$ such that the corresponding model estimate provides a good approximation to $f$, where quality of the estimates can be measured via some loss function $\mathcal{L}(\hat{f}(X),Y)$ or predictive risk $R(\hat{f}) = \mathbb{E}(\mathcal{L}(\hat{f}(X),Y) \, | \, \hat{f})$.  Research on model selection has traditionally focused heavily on assessing consistency and establishing conditions under which it can be guaranteed that the estimated $\hat{\lambda}_n$ converges to the true or optimal value (see e.g.\ \cite{Zhao2006,Bach2008,Zhang2009} or \cite{Fan2010} for a thorough overview). 
 
Such a selection process, however, says nothing about the stability or optimal-uniqueness of the particular (single) model ultimately selected.  Indeed, while it's quite rare that applied scientific publications provide point estimates \emph{without} an accompanying measure of uncertainty (e.g.\ confidence interval or an estimate of the standard error), it's at least as rare that selected models themselves come \emph{with} analogous measures (e.g.\ a collection of plausible models $\mathcal{M}^* \subseteq \mathcal{M}$).  Instead, readers are left to wonder, for example, whether some of the selected covariates could be substituted for others without a significant drop in risk or, more generally, if the entire procedure were repeated on a new sample, how different a model overall might be expected.  Such questions are particularly important in high-dimensional settings and/or when complexity restrictions (e.g.\ the number of covariates in a model) are enforced for practical purposes. 

The fact that in any given data modeling context there may exist many near-optimal models was a point stressed by Leo Breiman in his 2001 ``Two Cultures" essay \citep{Breiman2001} where he refers to the phenomenon as the ``Rashomon Effect''.  In the last several years, the idea of Rashomon or uncertainty \emph{sets} has emerged as a way of formally referring to sets of models with errors that are within some small $\epsilon$ of that of the empirically optimal model.  Within the recent Rashomon literature are methods for decision making \citep{Tulabandhula2013, Tulabandhula2014}, identifying ranges of variable importance within a model class \citep{Fisher2019}, and measuring model class simplicity \citep{Semenova2019}. 

\subsection{Model Set Selection}
While a number of interesting analyses can be carried out once a set of similarly-accurate models is identified, finding an efficient means of obtaining these model sets has proven more of a challenge.  \cite{Jiang2008} proposed a ``fence" method to weed out poor-performing models from $\mathcal{M}$ in hopes of obtaining some reduced set $\mathcal{M}^*$ that contains the optimal model with high probability.  \cite{Hansen2011} followed a similar regime, focusing primarily on linear modeling and proposing a sequential elimination strategy.  \cite{Ferrari2015} extended these ideas to potentially high-dimensional settings where $p$ can grow with $n$ whenever additional screening methods are available.  \cite{Li2018} developed the notion of ``model confidence bounds" for linear models chosen via penalized likelihood selection methods whereby a sequence of nested models is obtained in hopes that the true model lies in between the smallest and largest such models. \cite{Lei2020} developed a procedure for quantifying the uncertainty associated with models selected via sample-splitting and cross-validation, providing some means of determining $\mathcal{M}^*$ without insisting on overly strong conditions on the data and models.

Despite these impressive efforts, there remain a substantial number of key shortcomings.  With the exception of \cite{Lei2020}, these procedures are generally specific to a particular modeling framework -- often linear regression -- and their validity and theoretical properties depend heavily on that particular model class assumed.  Additionally, as pointed out in \cite{Li2018}, earlier procedures lack any explicit means of restricting the model structure and thus models in the selected set $\mathcal{M}^*$ needn't be related in any way and thus could potentially consist of completely disjoint sets of covariates.  These procedures also rely heavily on exhaustive search or backward elimination-type procedures, making them computationally daunting and potentially infeasible in high-dimensional settings without a valid variable screening tool \citep{Lei2020,Jiang2008}.  Finally, as pointed out in \cite{Lei2020}, without \emph{a priori} size or complexity restrictions, the procedures in \cite{Hansen2011}, \cite{Ferrari2015}, and \cite{Jiang2008} will always produce a set of models $\mathcal{M}^*$ that contains the fully saturated model with all available covariates.  

Our work here proposes a novel procedure for producing plausible model sets in a computationally efficient manner while both enforcing a notion of structural similarity and while remaining entirely agnostic to the particular models and loss functions employed.  Indeed, our model path selection (MPS) tool should be seen as a wrapper that can be applied to any user-specified model and loss function.  Instead of sequentially \emph{eliminating} models, MPS \emph{builds up} a plausible model set $\mathcal{M}^*$ in a manner akin to forward selection.  When low-dimensional models are desired, the final output is a collection of similarly predictively-optimal models that can be displayed in an intuitive, path-style graphical fashion that makes clear which covariates and portions of the model can be swapped for others without a substantial drop in accuracy.  When few paths are discovered, practitioners may take additional confidence in the potential importance of the included covariates while the appearance of numerous paths should discourage the over-emphasis of particular covariates that appear in only a small fraction of the models.  In the simulations and real-data applications provided in later sections, we repeatedly demonstrate that our MPS procedure is capable of generating relatively small collections of models, a surprising number of which have better out-of-sample accuracy than the single models identified via traditional tools like lasso and forward selection.

Before proceeding, it is worth pausing to distinguish our goal at hand from others that may appear similar and where substantially more attention has traditionally been paid.  We do not necessarily assume that the set of predictors can be partitioned into signal and noise covariates and thus our goal is not necessarily to identify the subset of signals as is the primary objective with tools like knockoffs \citep{Barber2015,Candes2018,Barber2018,Janson2016}, pseudovariables \citep{Wu2007,Hu2018}, or stability selection (SS) \citep{Meinshausen2010,Shah2013}.  We emphasize the distinction from SS in particular because our MPS procedure makes use of similar ideas involving the evaluation of model or variable selection frequencies across resamples.  Finally, we are not seeking to select a particular model and adjust variance estimates to account for additional uncertainty as might be done via some form of post-selection inference (e.g.\ \cite{Berk2013,Lee2016,Tibshirani2016,Taylor2018}).  Instead, our procedures are designed to capture and display the uncertainty involved in the model selection process itself and produce small sets of plausible models as output.

The remainder of this paper is laid out as follows.  In Section \ref{sec:motivation} we provide a high-level motivation for the proposed model path selection framework by building on core ideas from the stability selection (SS) literature.  In particular, we introduce the idea of \emph{forward stability selection} (FSS) as something of a bridge between SS and model path selection (MPS).  Section \ref{sec:methods} formalizes the FSS and MPS ideas, drawing upon classic results from the ranking and selection literature.  Numerous simulations and applications are provided in Sections \ref{sec:simulations} and \ref{sec:applications} before concluding with a discussion in Section \ref{sec:discussion}.  An R package containing the MPS procedure and code needed to reproduce our simulations and real data examples is available at \url{https://github.com/nkissel/MPS}.

\section{Background and Motivation}
\label{sec:motivation}

As above, assume we have a dataset $\mathcal{D}_n$ consisting of data $Z_1, ..., Z_n \sim F_Z$ in the form of ordered pairs $Z=(X,Y)$ consisting of covariates $X = (X_1, ..., X_p)$ and a response $Y$ and imagine a generic regression relationship of the form $Y = f(X) + \epsilon$.  In a variable selection context, one often further assumes that $X$ can be partitioned into $(S,N)$, where $S$ consists of $s$ signal variables and $N$ contains only noise features, defined as those that are independent of $Y$, at least conditional on the covariates in $S$.  

Stability selection (SS) is one popular resampling-based method for estimating $S$ that was introduced by \citet{Meinshausen2010} and has since been refined by \citet{Shah2013}. The basic SS procedure operates as follows. First, $B$ resamples $\mathcal{D}_1^*,\dots,\mathcal{D}_B^*$, each of size $\lfloor \frac{n}{2} \rfloor$, are drawn without replacement. On each resample $\mathcal{D}_{i}^{*}$, a collection of models $ \{ M_{\lambda^\prime}(\mathcal{D}_{i}^{*}):\lambda^\prime \in \Lambda^\prime \} $ are fit across a grid of tuning parameters $ \lambda^\prime \in \Lambda^\prime \subset \Lambda $.  Each model generated on each resample produces a set of selected covariates $\hat S_i(\lambda^\prime)$. For each tuning value $\lambda^\prime$ we then calculate the frequency of selection for each covariate across all resamples so that for each covariate $X_j$, the corresponding selection proportion is defined as 
\[
\hat\theta_j(\lambda^\prime)=\frac{1}{B}\sum_{i=1}^{B} { \mathbb{I}{\{X_j \in \hat S_i(\lambda^\prime)\}} }.
\]
To form the final estimated set of signal covariates $\hat S^{stable}$, one selects an appropriate tuning parameter value $\lambda_0$ and threshold $ \pi_{thr} \in [0,1]$ so that $\hat S^{stable}$ can then be defined as the collection of all covariates $X_j$ such that $\hat \theta_j(\lambda_0) > \pi_{thr}$.  Work following the inception of stability selection (e.g. \cite{Shah2013}) has focused on establishing assumptions under which particular error controls can be achieved so as to better guide the practical selection of appropriate choices for $\pi_{thr}$ and $\lambda_0$. 

At a high level, SS can be thought of as a resampling-based variable selection procedure that takes a modeling framework and data as inputs and outputs a set of selected covariates $\hat S^{stable}$.  Crucially, this setup leaves one with little ability to directly control the complexity, structure, or size $\hat{s}$ of $\hat S^{stable}$.  Especially in the modern era, however, there exists many situations in which hundreds or thousands of covariates are available, a substantial proportion of which may contain some signal, but where, for practical purposes, users desire a low-dimensional model consisting of only a handful of variables. 

Suppose, for example, a practitioner desires a model with no more than $k_{max}$ covariates.  The simplest way to accommodate this restriction within the SS framework in the event that $k_{max} < \hat{s}$ would be to simply select those $k_{max}$ covariates with the highest empirical selection proportions.  This, however, says nothing with regard to how uniquely optimal those particular variables may be relative to the others in $\hat S^{stable}$ and perhaps even more importantly, there is no reason to expect that the model constructed with those particular $k_{max}$ covariates with the highest selection proportion would be the most predictively accurate $k_{max}$-variable model.  

\begin{figure}[t]
\includegraphics[width=1\columnwidth]{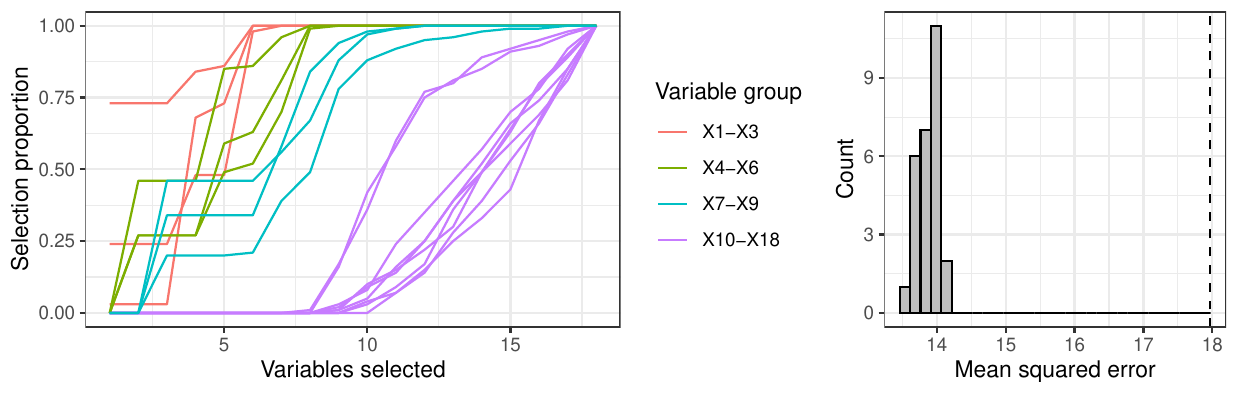}
\caption{(Left) Stability paths from the motivating model in (\ref{eqn:motivation}). Stability selection is performed using forward selection. (Right)  Histogram of test set MSEs from all 3-variable models that take one variable from each signal grouping. The dashed vertical line shows the best 3-variable model built by including any 3 variables from the first 2 signal groupings. \label{fig:motive}}
\end{figure}

As a simple motivating example, consider the regression function 
\begin{equation}
\label{eqn:motivation}
\mathbb{E}[Y|X] = 3 \sum_{i=1}^{3} X_i \, + \, 2 \sum_{j=4}^{6} X_j \, + \, 1 \sum_{k=7}^{9} X_k \, + \,  0 \sum_{\ell=10}^{18} X_\ell
\end{equation}
where each of the four groups of covariates are independent but within the first three signal groups, the covariates have a high correlation of 0.9.  In this example, one might expect that among the $\binom{18}{3}$ possible 3-variable linear models, the most accurate 3-variable models would include one covariate from each signal group:  one variable from $\{ X_1,X_2,X_3 \}$, one from $\{ X_4,X_5,X_6 \}$, and one from $\{ X_7,X_8,X_9 \}$.  If, however, we naively employ stability selection for this purpose, the three covariates with the highest selection proportions are very often from the first 2 groupings.  Stability paths from this procedure are displayed on the left-hand side of Figure \ref{fig:motive}. Here stability selection is implemented with forward selection on a dataset with $n=500$ observations and a signal-to-noise ratio (SNR) of approximately 10, and model performance is measured by the MSE on a test dataset of 10,000 observations.

As suspected, however, such models are far from the most accurate 3-variable models available.  The right-hand side of Figure \ref{fig:motive} shows a histogram of MSEs from all 3-variable models that take one variable from each signal grouping.  The dashed vertical line corresponds to the error of the \emph{best} 3-variable model built by including any three variables from only the first two signal groupings, thereby serving as a lower bound on the error for any model obtained via this stability selection approach on this dataset.  It is immediately evident looking at this plot that even the least accurate models formed by taking one covariate per group perform substantially better than even the best model obtained via the SS approach.  

How else then might we embed the idea of stability within the context of model selection?  Here we focus on the generic idea of stepwise forward selection whereby covariates enter the model one at a time and at each step, the covariate selected is that which minimizes the empirical loss of the resulting model. In the original SS formulation, \cite{Meinshausen2010} suggested that models built via forward selection be constructed on each resample to achieve a notion of stable \emph{variable} selection.  Here, in order to achieve stable \emph{model} selection, we advocate for the stability achieved via resampling to be embedded within the model fitting procedure itself.  More specifically, in the context of forward selection, suppose that at each step we draw $B$ resamples and determine the best variable to add based on each resample.  At each step, the end result is a selection proportion for each remaining covariate and we can simply choose to add the covariate selected most often across those resamples.  We refer to this idea as \emph{forward stability selection}.  

The formulation of such a procedure begs an interesting question, however.  Suppose that at any given step in the procedure, $k>1$ covariates are selected quite frequently and in particular, none of those $k$ appear to be selected significantly less often than the other top performers.  In such cases, rather than select only the one covariate with the empirically largest selection proportion, our \emph{model path selection} (MPS) procedure, formalized in the following section, involves creating a new model path for each of these $k$ covariates and continuing the model-building procedure for each, stopping when some $k_{max}$ number of covariates is reached in each model path.  In this way, MPS generates a collection of models rather than only a single one, while remaining entirely agnostic to the particular model class and loss function desired.


\section{Methods}
\label{sec:methods} 

Considering the generic regression framework $Y = f(X) + \epsilon$, model set selection (MSS) procedures seek to identify the best estimators of $f$ among a finite set of candidate models $\mathcal{M} = \{\hat{f}_m: m=1, ..., M \}$.  As recently discussed in \cite{Lei2020}, the MSS objective can be posed as a hypothesis testing problem with 
\begin{align}
\label{eqn:hyp}
H_{0, m}: \; Q(\hat f_m) &\leq  Q(\hat f_{m^\prime})  \text{ for all } m^\prime \neq m \\ 
H_{1, m}: \; Q(\hat f_m) &>  Q(\hat f_{m^\prime}) \text{ for some } m^\prime \neq m \nonumber
\end{align}
where $Q(\hat f) = \E [\mathcal{L}(\hat f(X), Y )| \hat f]$ denotes the predictive risk with respect to a loss function $\mathcal{L}$.  In this setup, failing to reject $H_{0,m}$ means that we do not have evidence that any other candidate model in $\mathcal{M}$ generates more accurate predictions than $\hat{f}_m$, and as a result, the model $\hat{f}_m$ is added to the selected set $\mathcal{M}^* = \{\hat{f}_m: \text{fail to reject } H_{0,m} \} \subseteq \mathcal{M}$. 

Formulated in this way, it becomes apparent why MSS methods generally tend to require an exhaustive search over the entire set of candidate models and as a result, are computationally expensive.  Indeed, this MSS setup is exactly analogous to best subset selection in the traditional model selection paradigm.  In those settings, however, there are a multitude of alternative selection strategies that perform more greedy searches in lieu of exhaustive searches in order to reduce computational cost. Specifically, consider a generic forward stepwise regression which, at each step, selects the covariate that minimizes the error of the model given the covariates already selected. Stepwise methods terminate when some pre-specified stopping criterion is met, of which there are many possibilities.  Here, for simplicity, we focus our attention on the simplest variant that stops once a pre-specified number of covariates $d$ are selected. Greedy selection allows one to quickly find an accurate $d$-covariate model by selectively searching over $\sum_{j=1}^{d} (p-j+1)$ models rather than all possible $\binom{p}{d}$ $d$-covariate models. For scenarios with large $p$ and $1 < d < p$, the computational gains of forward selection procedures can be immense.  The goal of our model path selection (MPS) routine is thus to provide an analogous, computationally efficient alternative to exhaustive search MSS procedures that still takes into account the uncertainty in the selection process by identifying an entire subset of accurate models $\mathcal{M}^* \subseteq \mathcal{M}$.

\begin{algorithm}[t]
\SetKw{KwIn}{in}
Select desired model size (number of covariates) $d$\;
Define $\mathcal{M}^*_0$ as a set only containing the null model\;
\For{$i \gets 1$ \KwTo $d$ }{
 \For{$\hat{f}_m$ \KwIn $\mathcal{M}_{i-1}^*$ }{
  Define $\mathcal{M}_i(m)$ as the set of all  $i$-variable models that nest $\hat{f}_m$\;
  Perform model set selection (MSS) on $\mathcal{M}_i(m)$ and call the result $\mathcal{M}_i^*(m)$\;
 }
 Set $\mathcal{M}_i^* = \bigcup_{\hat{f}_m \in \mathcal{M}_{i-1}^*} \mathcal{M}_i^*(m)$
 
}
Set $\mathcal{M}^* = \mathcal{M}^*_d$\;
\caption{\label{alg:mps}
(Generalized) Model path selection (MPS)}
\end{algorithm}

As in traditional forward selection, the generic MPS procedure begins by comparing all $p$ univariate models, which we denote by $\mathcal{M}_1$.  We then perform MSS on $\mathcal{M}_1$ to generate the subset of the most accurate 1-variable models, denoted by $\mathcal{M}_1^*$. Then, for each model $\hat{f}_m \in \mathcal{M}_1^*$, let $\mathcal{M}_2(m)$ denote the set of all 2-covariate models nesting $\hat{f}_m$ and perform MSS on this set to generate $\mathcal{M}_2^*(m)$.  We can then define $\mathcal{M}_2^* = \bigcup_{\hat{f}_m \in \mathcal{M}_1^*} \mathcal{M}_2^*(m)$ as the set containing all of the most accurate 2-covariate models that nest one of the 1-variable models in $\mathcal{M}_{1}^{*}$.  More generally, let $\mathcal{M}_j(m)$ represent all $j$-variable models that nest some $\hat{f}_m \in \mathcal{M}^*_{j-1}$ for some $j \leq d$.  MSS is then performed yielding the selected set $\mathcal{M}^*_j(m) \subseteq\mathcal{M}_j(m)$ and we define 
$$
\mathcal{M}^*_j = \bigcup_{\hat{f}_m \in \mathcal{M}^*_{j-1}} \mathcal{M}^*_{j}(m).
$$ 
When $j=d$, the procedure terminates and the final selected set $\mathcal{M}^*$ is set equal to $\mathcal{M}_d^*$.  This MPS procedure is summarized in Algorithm \ref{alg:mps}.

Note that the MPS procedure defined in Algorithm \ref{alg:mps} is intentionally given in a generalized form.  In particular, so long as the variable selection is carried out in a forward fashion, any MSS procedure can be employed to determine the selected covariates at each step.  The output of the MPS procedure is thus not only a collection of predictively accurate models, but a collection that, by construction, remain structurally similar.  The models selected can be easily displayed in a natural tree-like branching fashion, which makes the output immediately and naturally interpretable.  Much more discussion along these lines is provided in Subsection \ref{sec:graphics}; readers are invited to look ahead to this future subsection and inspect Figure \ref{fig:ex1} to see example output.

\subsection{Stabilized Selection}

As just noted above, MPS can be carried out with any form of MSS one desires.  Note, however, that MPS insists on a kind of stepwise MSS.  That is, along each model path, at each step $j$, we perform a constrained version of MSS where the collection of models identified must include the $j-1$ covariates selected previously in that path.  Thus, in essence, we are seeking only to identify the subset of remaining covariates most helpful to be added to the model already containing those $j-1$.  Thus, as alluded to in Section \ref{sec:motivation}, we now propose a new kind of MSS building on the idea of stability and more specifically tailored for this type of situation.

Consider a particular model $\hat{f}_{(j-1)} \in \mathcal{M}^{*}_{j-1}$ that, without loss of generality, contains the first $j-1$ covariates so that we may write $\bm{X}_{j-} = \{X_1, ..., X_{j-1} \}$ as the set of covariates already selected and $\bm{X}_{j+} = \{X_j, ..., X_{p} \}$ as the set of covariates available to be added.  Given our original dataset $\mathcal{D}_n$, at the $j^{th}$ step in the procedure, there must exist some (at least one) covariate in $\bm{X}_{j+}$ that minimizes the loss of the resulting model when added.  Recall, however, that the entire motivation for MSS procedures and the fundamental idea of model stability is that given a different dataset, a different model may appear empirically optimal.  Thus, by exactly the same reasoning, if we had obtained a different dataset, a different covariate may appear optimal at step $j$.

To account for this uncertainty in the selection process, suppose we draw $B$ resamples of the original dataset $\mathcal{D}^{*}_{1}, ..., \mathcal{D}^{*}_{B}$ and obtain the empirically optimal covariate to be added relative to each resample.  For each resample, define the count variable for the $k^{th}$ covariate $C_{k,i}$ to be equal to 1 if $X_k$ is selected on the $i^{th}$ resample and equal to 0 otherwise.  We can then define the empirical selection proportion
\[
\hat{\theta}_k = \frac{1}{B} \sum_{i=1}^{B} C_{k,i}
\]
to measure the frequency with which $X_k$ was selected and do the same for all remaining covariates.

Formulated in this fashion, we can simplify the generic MSS task -- rather than select every model not significantly worse than any other, at each step in MPS, we need only select all covariates not selected significantly less often than any other.  Thus, in our MPS context, at each step $j$, the hypotheses in (\ref{eqn:hyp}) can be rewritten as 
\begin{align}
\label{eqn:newhyp}
H_{0, k}^{(j)}: \; \theta_k &\geq \theta_q \text{ for all } X_q \in \bm{X}_{j+} \\
H_{1, k}^{(j)}: \; \theta_k &< \theta_q \text{ for some } X_q \in \bm{X}_{j+}. \nonumber
\end{align}

Failing to reject $H_{0, k}^{(j)}$ means that we do not have sufficient evidence to conclude that $X_k$ is not one of the most selected covariates.  As a result, the model containing $X_k$ along with the previously selected $j-1$ covariates is added to the selection set $\mathcal{M}_{j}^{*}$ at step $j$.

It's worth pausing to stress a few points.  First, in the interest of clarification, note that at each step $j$ in the MPS procedure, this identification of a subset of suitable covariates must be carried out for each model in $\mathcal{M}_{j-1}^{*}$ obtained to that point.  Furthermore, we stress that while the reformulation of the problem above is intuitively convenient, obtaining such a subset remains quite a nontrivial problem.  Depending on the data, model, and loss function employed, there may be many equally suitable covariates, none of which have a high selection frequency.  Rather, both the maximum selection proportion and the number of covariates without a significantly lower selection proportion are unknown for each model at each step.  For help in solving this problem, we turn to a classical set of literature on the problem of \emph{ranking and selection}.

\subsubsection{Ranking and Selection}
As just outlined, the procedure described above involves generating $B$ resamples of the original data for each model at each step of the MPS procedure and determining the subset of covariates selected most often across those resamples.  Imagine now an idealized setting in which rather than resampling, those datasets $\mathcal{D}_{1}^{*}, ..., \mathcal{D}_{B}^{*}$ are instead independently generated datasets of size $n$ sampled directly from the population.  In this scenario, the selection of a single covariate based on each dataset could equivalently be thought of as a multinomial sample of size 1.  That is, we can think of this process as producing $B$ multinomial samples, each of the form $W_i = (0, ..., 0, 1, 0, ...0)$ where the entry at index $k$ is set to 1 whenever $X_k$ is selected on the $i^{th}$ dataset and all remaining entries are set to 0.  In addition to $B$, this multinomial distribution is thus parameterized by the true selection probabilities $(\theta_j, ..., \theta_p)$ whenever $X_j, ..., X_p$ are assumed to be the remaining covariates available at step $j$.  This problem of selecting a subset of covariates can thus be transformed into the problem of identifying the most probable categories in a multinomial. We can therefore leverage the multinomial dependence between all $\theta_i$ to generate a covariate selection tool, rather than explicitly considering the collection of hypotheses in (\ref{eqn:newhyp}).

Though first applied to ranking normal populations, the ranking and selection literature quickly moved to the ranking and selection of multinomial parameters from the same population \citep{bechhofer1959,Kesten1959}, focusing primarily on exactly this problem -- identifying the most probable multinomial event(s) \citep{bechhofer1959,Alam1971,Gupta1967}.  Several of those methods proposed are immediately applicable to the selection problem described above \citep{Gupta1967, Panchapakesan1971}. 

While much of the early literature focused on bounding the probability that the event with the largest empirical count corresponded to the most probable event, the methods developed in \citet{Gupta1967} are the first to our knowledge that identify a set of multinomial cells in such a way that bounds the probability that the most probable event(s) are contained in the selected set.  Various other papers have followed suit, creating their own subset selection procedures or further refining existing methods \citep{Panchapakesan1971, Chen1986, Bechhoffer1988, Chen1989}.

In order to obtain such a probabilistic bound we utilize the following decision rule defined in \citet{Panchapakesan1971}:

\bigskip

\noindent R: \textit{Continually sample from multinomial $\Theta$ until one cell reaches a frequency of $r$, then stop sampling and select all cells with frequencies at least $r - D$.}

\bigskip

\noindent  Within this framework, it has been proven that the probability of making a correct selection -- denoted $\mathbb{P}(C \, | \, r,D)$ -- for \emph{any} multinomial, is bounded by the probability of making a correct selection when the true multinomial distribution has uniform probability parameters; that is, $\theta_m = \frac{1}{M}$ for all $m = 1, ..., M$ in a multinomial with $M$ possible outcomes \citep{Chen1986,Liu1991, Panchapakesan2006}.

Note that this selection procedure deviates slightly from the fashion in which the covariate subset selection problem was discussed above.  In particular, we specify the maximal cell count $r$ rather than the total number of new datasets $B$.  The total number of new datasets needed is, however, still bounded above by $M(r-1)+1$ whenever there are $M$ covariates still available to be selected.  Note also that in applying the decision rule $R$, the choice of $D$ and $r$ can be user-specified so as to induce a particular lower bound on the probability $P^*$.  Of course, in our context, we wish to find the smallest possible set for which we have a minimal user-specified probabilistic bound $P^*$.  We therefore specify a desired maximal cell count $r$ and minimum probability $P^*$ and solve for the smallest $D$ for which $\mathbb{P}(C \, | \, r,D) \geq P^*$ can be guaranteed to hold.  This task can be easily accomplished via simulation; an empirical solution is given in Algorithm \ref{alg:findwindow}.

\begin{algorithm}[t]
\KwResult{Smallest allowable value of $D$}
Set desired number of simulations $nsim$\;
\For{$h \gets 1$ \KwTo $nsim$}{
 Set $x^{(h)}_m \gets 0$ for all $m \in \{ 1,\dots,M \}$\;
 \While{$\max_m(x^{(h)}_m) < r - D$}{
  Take sample from uniform multinomial $\Theta(\frac{1}{M}, \dots, \frac{1}{M})$\;
  Record the sampled cell's index as $i$\;
  Set $x_{i}^{(h)} = x_{i}^{(h)} + 1$
 }
}
 Across all $h$, find the smallest value of $D$ for which $\#(x_1^{(h)} \geq r - D) \geq \lceil nsim * P^*\rceil$\;
 \textit{Note}: $x_1^{(h)}$ is chosen WLOG--any fixed subscript can be used\;
 \caption{\label{alg:findwindow}
 Empirically finding $D$ to guarantee $P^*$}
\end{algorithm}

\subsubsection{Resampling}

In practice, certainly it is not reasonable to expect to have $B$ newly and independently generated datasets from the population available at each step in the MPS procedure, but we can approximate this process by resampling from the original dataset $\mathcal{D}_n$.  In deciding between bootstrapping and subsampling, note that by seeking to identify the most probable covariates to include, we are dealing with a discrete-valued parameter, which is known to be a common source of bootstrap failure \citep{Bickel1997,Davison1999}.  On the other hand, in order for subsampling to work asymptotically, we need only to subsample at a rate of $o(n)$ and that a standardized version of our statistic converges to a non-degenerate distribution \citep{Politis1999}.

As a simple demonstration of the shortcomings of the bootstrap in this context, consider the simple additive model $Y=X_1+X_2+\epsilon$ where the covariates and noise are all independent and each is sampled in an iid fashion from a standard normal distribution.  Suppose further that our complete set of candidate models $\mathcal{M}$ contains only the two univariate linear models.

In this simple scenario, it's clear that we should expect each model to be selected equally often -- since the covariates are independent, from the same distribution, and have the same relationship to the response. For any given dataset, the univariate model selected is simply that with the covariate that appears (by random chance) to be more strongly correlated with $Y$.  Thus, averaging across datasets, we should expect that the proportion of times each model is selected to be roughly 0.5.

\begin{figure}[]
	\centering
	\includegraphics[width=1\textwidth]{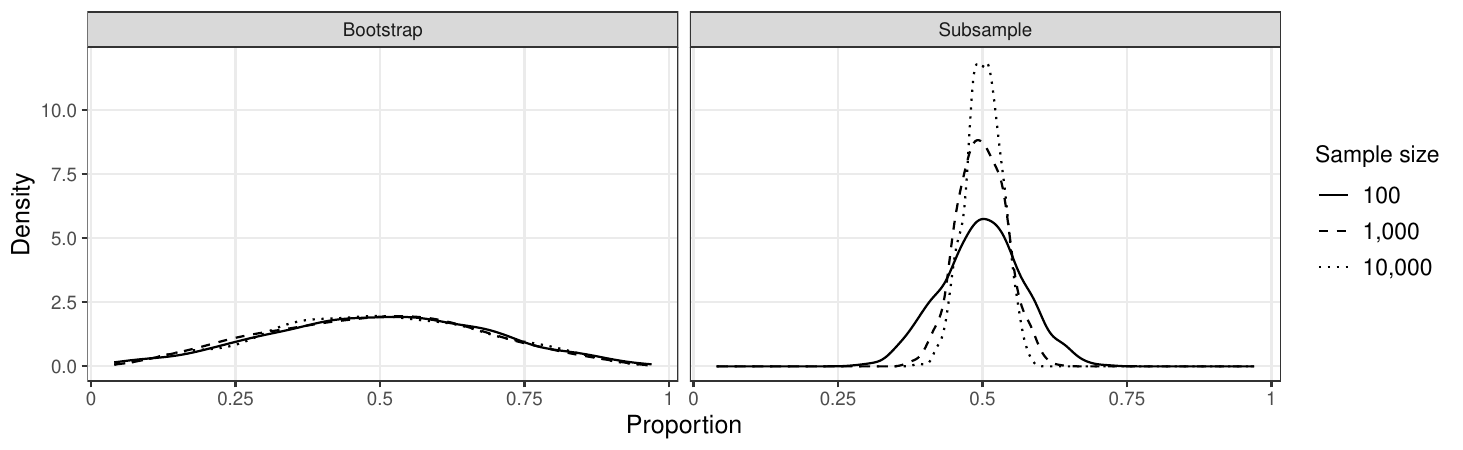}
	\caption{Selection proportion distribution of $X_1$ when bootstrapping vs subsampling on the order of $\sqrt{n}$.}
	\label{fig:resamp}
\end{figure}

Figure \ref{fig:resamp} looks at the distribution of these selection proportions whenever those datasets are resampled from a single original sample via either bootstrapping or subsampling.  We generate a datasets with $n = 100, 1000, $ and 10,000 observations, generate $B=500$ resamples of each, and calculate the proportion of times across those resamples that $X_1$ is selected as the more important covariate.  The entire procedure is repeated 1000 times at each sample size to generate distributions of these selection proportions at the three different sample sizes.  When subsampling at a rate of $\sqrt{n}$, these distributions behave as expected -- each is centered at approximately 0.5 with higher concentration for larger $n$.  With bootstrapping, however, each distribution is nearly identical across all $n$ with a noticeably larger variance.  In implementing the model set selection procedure via the ranking and selection approach outlined above, we thus recommend that the resampling for each model at each step be carried out via subsampling at a rate slower than $n$.

\subsubsection{Summarizing the MPS Procedure}

\begin{algorithm}[t]
\SetKw{KwIn}{in}
Select desired model class, loss $\mathcal{L}$, and size (number of covariates) $d$\;
Select maximum cell count $r$ and minimum probability $P^{*}$\;
Define $\mathcal{M}^*_0$ as a set only containing the null model\;
\For{$i \gets 1$ \KwTo $d$ }{
 \For{$\hat{f}_m$ \KwIn $\mathcal{M}_{i-1}^*$ }{
 	Denote the remaining covariates by $X_{[1]}, ..., X_{[p-i+1]}$\;
 	Set $C_{k} = 0$ for $k=1 ,..., p-i+1$\;
 	\While{$\max_k(C_k) < r$}{
	Generate subsample $\mathcal{D}^{*}$ of size $\sqrt{n}$\;
	Select optimal covariate $X_{[k^*]}$ that minimizes $\mathcal{L}$ on $\mathcal{D}^{*}$\;
	Set $C_{k^*} = C_{k^*} + 1$\;
 }
  Run Algorithm \ref{alg:findwindow} with $r$ and $P^{*}$ to obtain minimum value $D$\;
  \For{$k \gets 1$ \KwTo $p-i+1$ }{
  Let $\hat{f}_{m,[k]}$ denote the model that includes  $X_{[k]}$ and the covariates in $\hat{f}_m$\;
  If $C_k > r - D$, include $\hat{f}_{m,[k]}$ in $\mathcal{M}_i^*(m)$\;
  }
 }
 Set $\mathcal{M}_i^* = \bigcup_{\hat{f}_m \in \mathcal{M}_{i-1}^*} \mathcal{M}_i^*(m)$
 
}
Set $\mathcal{M}^* = \mathcal{M}^*_d$\;
\caption{\label{alg:mps2}
Model path selection (MPS)}
\end{algorithm}

With the above issues addressed, we can now present a final summary of our proposed model path selection (MPS) procedure that involves resampling and making use of the ranking and selection strategy outlined above to perform the model set selection (MSS) for each model at each step.  This MPS procedure is summarized in Algorithm \ref{alg:mps2}.  Note that as discussed in previous sections, if we desired only a single model output but still wanted to incorporate stability into the selection procedure, rather than performing a MSS at each step, we could instead  simply take $B$ subsamples at each step and keep only that covariate with the largest empirical selection frequency.  The end result is merely a stabilized form of forward stepwise selection and we thus refer to it as \emph{forward stability selection}.

While this MPS formulation outlined in Algorithm \ref{alg:mps2} is, by construction, a specialized form of the generalized MPS presented at the beginning of Section \ref{sec:methods}, this is the form of MPS that will be utilized in the remainder of the paper unless otherwise specified.  The simulation results presented in the following section provide strong empirical evidence that this version of MPS substantially outperforms alternative versions in terms of both (i) its computational efficiency and (ii) its ability to select accurate models while maintaining a relatively small collection of such models. 

Note that as alluded to above, in addition to specifying the model class and loss function, this formulation also requires the user to specify a maximum cell count $r$ and a minimum desired probability $P^*$.  The $r$ parameter is directly analogous to the number of resamples one would employ in similar procedures and thus should be set to some large positive value subject to computational constraints.  The $P^*$ parameter, which corresponds to the probability that the selected set of covariates contains that with the true maximum selection frequency, can be thought of in the same fashion as a confidence level. We would generally expect most users to fix this at a standard default value; for example, 0.90 or 0.95.  Note that just as constructing confidence intervals with larger confidence levels results in wider intervals, higher values of $P^*$ mean that more covariates are likely to be selected at each step, leading to ``wider" path outputs with more models selected.

\subsection{Graphical Visualization}
\label{sec:graphics}

Due to its branching nature, the path structure output of MPS can easily be displayed as a collection of trees, wherein each tree corresponds to a covariate selected in the first step.  In each tree, the depth describes the iteration at which a covariate was selected and sibling nodes represent covariates that were deemed similarly predictively useful conditional on the model produced by selecting the covariates in the parent nodes.  As a point of convention, these trees are left-justified, meaning the leftmost covariate is that with the highest empirical selection proportion and all sibling nodes are given in descending order of their empirical selection proportions. 

\begin{figure}[]
	\centering
	\includegraphics[width=.8\textwidth]{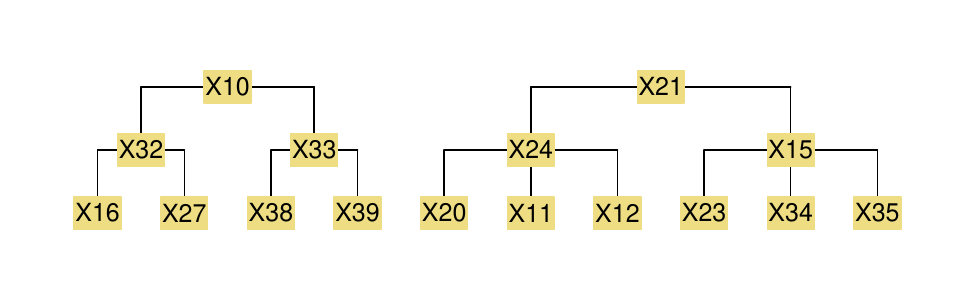}
	\caption{Example graphical output of the MPS procedure. }
	\label{fig:ex1}
\end{figure}

An example of the graphical output from the MPS procedure is shown in Figure \ref{fig:ex1}.  In this hypothetical example, imagine we sought to find a collection of similarly accurate 3-variable models given an initial collection of 40 covariates. We see that MPS identified a total of 10 such models spread across two separate trees.  At step one, two covariates, {X10} and {X21}, were selected, thereby forming the root nodes of the two trees.  At step two, both {X32} and {X33} were identified as potentially optimal covariates to be added to the model next, conditional on {X10} having been added first.  Similarly, {X24} and {X15} were selected in step two of the procedure when added to a model containing only {X21}. The pattern then continues for deeper depths where more covariates are added to the models identified in previous steps.

The fact that the output of the MPS procedure is so immediately and naturally interpretable provides a great deal of benefits to applied practitioners.  For example, even a brief glance at the collection of paths produced gives some intuition into the stability of the model selection task, with wider trees indicating that many accurate models are available. Further, the identification of numerous accurate models implies that emphasis needn't be placed on any single model produced via a classical model selection procedure like forward selection or lasso.  A demonstration of this is given on real datasets in Section \ref{sec:applications}.  It's also common in many applications for researchers to wonder whether, in a given chosen model, some covariates could be swapped for others with minimal additional loss.  The path-style output of MPS makes such questions immediately and visually answerable by simply checking whether such an alternative path exists.

Furthermore, this basic kind of plot shown in Figure \ref{fig:ex1} can be easily extended and modified to suit a variety of practical research restraints. Suppose, for example, each covariate has an associated price of collection. Researchers could then apply some sort of color gradient associated with price to easily display the ``cost" of each selected model.  A similar idea could be applied to missing data problems wherein, for example, researchers might wish to impute some missing covariate values prior to selecting the models.  Those covariates containing imputed values could then be made a different color in the MPS output.  Finally, in settings where large numbers of paths (models) are produced, a circular (radial) representation of the output may be more easily displayed.  An example of this alternative graphical style is shown in Figure \ref{fig:diapaths}.


\section{Simulations}
\label{sec:simulations}
We now present a number of simulations to investigate the performance and computational efficiency of the MPS procedure outlined above.  In order to allow for a robust set of comparisons, our focus here is on linear regression under the standard squared-error loss.  The particular setups described below very closely follow those in the recent work by \citet{Hastie2017} where the authors provide an in-depth comparison of forward selection, best subset selection, lasso, and relaxed lasso.

We consider data sampled from models of the form ${Y} = {X}{\beta} + \epsilon$ where ${X} \sim N(0, \Sigma)$ and where $\Sigma$ is a $p \times p$ Toeplitz matrix with row $i$ and column $j$ having covariance $\rho^{|i-j|}$. The noise term $\epsilon \sim N(0, \frac{1}{\nu} \beta^T\Sigma\beta)$ is a $n$-dimensional vector with $\nu$ enforcing a desired signal to noise ratio (SNR) and $\beta$ is a sparse $p$-dimensional vector with $s$ non-zero coordinates.  

Due to the Toeplitz structure of $\Sigma$, changing the sparsity pattern of $\beta$ has a non-trivial impact on the modeling environment. We therefore consider a variety of sparsity patterns in which we alter both the coordinates and magnitudes of the non-zero entries of $\beta$: 
\begin{itemize}
    \item \textbf{Beta-type 1}: $s$ non-zero components equal to 1 at approximately equally spaced positions from 1 to $p$;
    \item \textbf{Beta-type 2}: $s$ non-zero components equal to 1 in the first $s$ positions;
    \item \textbf{Beta-type 3}: $s$ non-zero components equally spaced from 10 to 0.5 in the first $s$ positions.
\end{itemize}

To carry out these simulations, we begin by fixing a number of settings including $n$, $p$, $s$, $\rho$, $\nu$ and the beta-type.  We then sample $X \in \tR^{n \times p}$ i.i.d. from $N(0, \Sigma)$ and $\epsilon \in \tR^n$ from $N(0,  \frac{1}{\nu} \beta^T\Sigma\beta)$ and calculate the response ${Y} = {X}{\beta} + \epsilon$.  With the dataset in hand, we fit a variety of models and perform a number of model set selection (MSS) procedures (described below), recording the performance of each.  We then repeat the entire procedure 112 times and record the average performance.

Performance is measured through a ratio of the test errors. To calculate this, we generate a test dataset of size  $10{,}000$ in the same manner as the training data and calculate the error of each regression estimate obtained. Our metric, the relative test error, is of the form
$$
\text{RTE}(\hat \beta) = \frac{\lVert Y_{test} - X_{test} \hat \beta \rVert_2^2}{\lVert Y_{test} -  X_{test} \beta \rVert_2^2}
$$
where $\lVert \cdot \rVert_2$ is the $L_2$ norm. Note that the provided RTE is slightly different from that presented in \citet{Hastie2017}, which opts for a ratio of the expected test errors.   Importantly, for MSS methods (in which multiple models are nearly always generated), we only report the \emph{minimum} RTE values from among the selected model set.  That is, only the best performing models are summarized in the RTE figures below.  We take this approach because our primary interest is in determining how often such methods identify \emph{some} models that outperform standard models like lasso and forward selection.

In each simulation setting, we investigate the following models and model set selection strategies: 
\begin{itemize}
\item \textbf{Oracle}: OLS is performed on only the $s$ covariates with non-zero coefficients.
\item \textbf{Stability Selection}: Stability selection is performed using forward selection and a final $s$-variable model is chosen by performing OLS on the $s$ most frequently selected variables.  This corresponds to the natural stability selection extension described in Section \ref{sec:motivation}.
\item \textbf{Lasso}: A lasso model with a regularization parameter chosen via a 10-fold cross validation on the training dataset across 100 possible values. 
\item \textbf{Forward Selection}: Forward stepwise linear regression where the covariate added at each step is that which minimizes the resulting squared error of the model.  We continue the process until the model contains $s$ covariates.
\item \textbf{Model Path Selection (MPS)}: MPS performed to a depth of $s$ using the resampling and ranking and selection approach outlined in Algorithm \ref{alg:mps2}.
\item \textbf{Cross-Validation with Confidence (CVC)}: An exhaustive search MSS method defined in \citet{Lei2020}. CVC uses cross validation, either $k$-fold or sample spitting, to form a MSS set by estimating model risk and assessing the hypotheses described in (\ref{eqn:hyp}). We perform 5-fold CVC on all $s$ variable models.
\item \textbf{MPS with CVC (CVC-MPS)}: MPS performed to a depth of $s$ using CVC as selection method for each model at each step. The CVCs performed as part of this procedure are also 5-fold.
\end{itemize}
CVC was chosen because in addition to being one of the most recently proposed MSS methods, it is also one of the most general,  imposing no structural restrictions on $\mathcal{M}$, which makes it easy to integrate within the MPS framework.  Note that all methods are performed so as to include only $s$-covariate models, except for lasso which is optimized via 10-fold cross validation as this is far more in line with its typical use in practice.

We consider three initial setups of $(n,p,s,r, P^*)$ where $r$ and $P^*$ represent the maximum cell count and inclusion probability, respectively, of the MPS procedure defined in Algorithm \ref{alg:mps2}:
\begin{itemize}
    \item \textbf{Setup 1:}  $n=100, p=10, s = 5, r = 200, P^* = 0.95$;
    \item \textbf{Setup 2:}  $n=500, p=100, s = 5, r = 200, P^* = 0.75$;
    \item \textbf{Setup 3:}  $n=500, p=100, s = 5, r = 50, P^* = 0.50$.
\end{itemize}
In each setting, we consider 5 values for the SNR ranging from 0.25 to 4 equally spaced on the log scale and autocorrelation levels $\rho$ of $0, 0.35$, and $0.7$. 

\begin{figure}[t!]
\includegraphics[width=0.84\columnwidth]{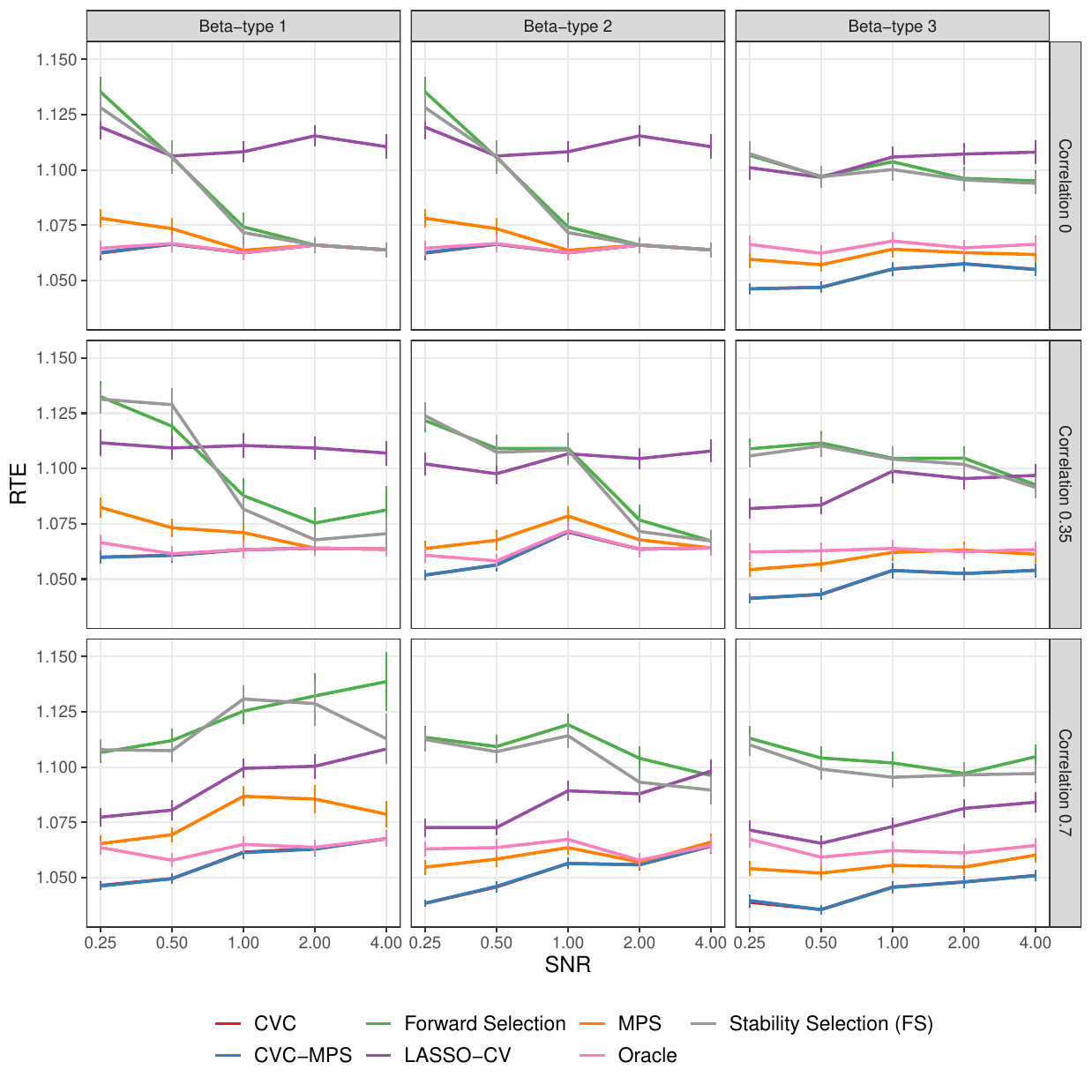}
\caption{Average RTE of each model and model set selection procedure.  The vertical bars represent 1 standard error. For each MPS, CVC, and CVC-MPS, the values represent the minimum RTEs from among the set of selected models.  Note that CVC and CVC-MPS perform nearly identically in each setting. 
\label{fig:sim10}}
\end{figure}

\begin{figure}[t!]
\includegraphics[width=0.84\columnwidth]{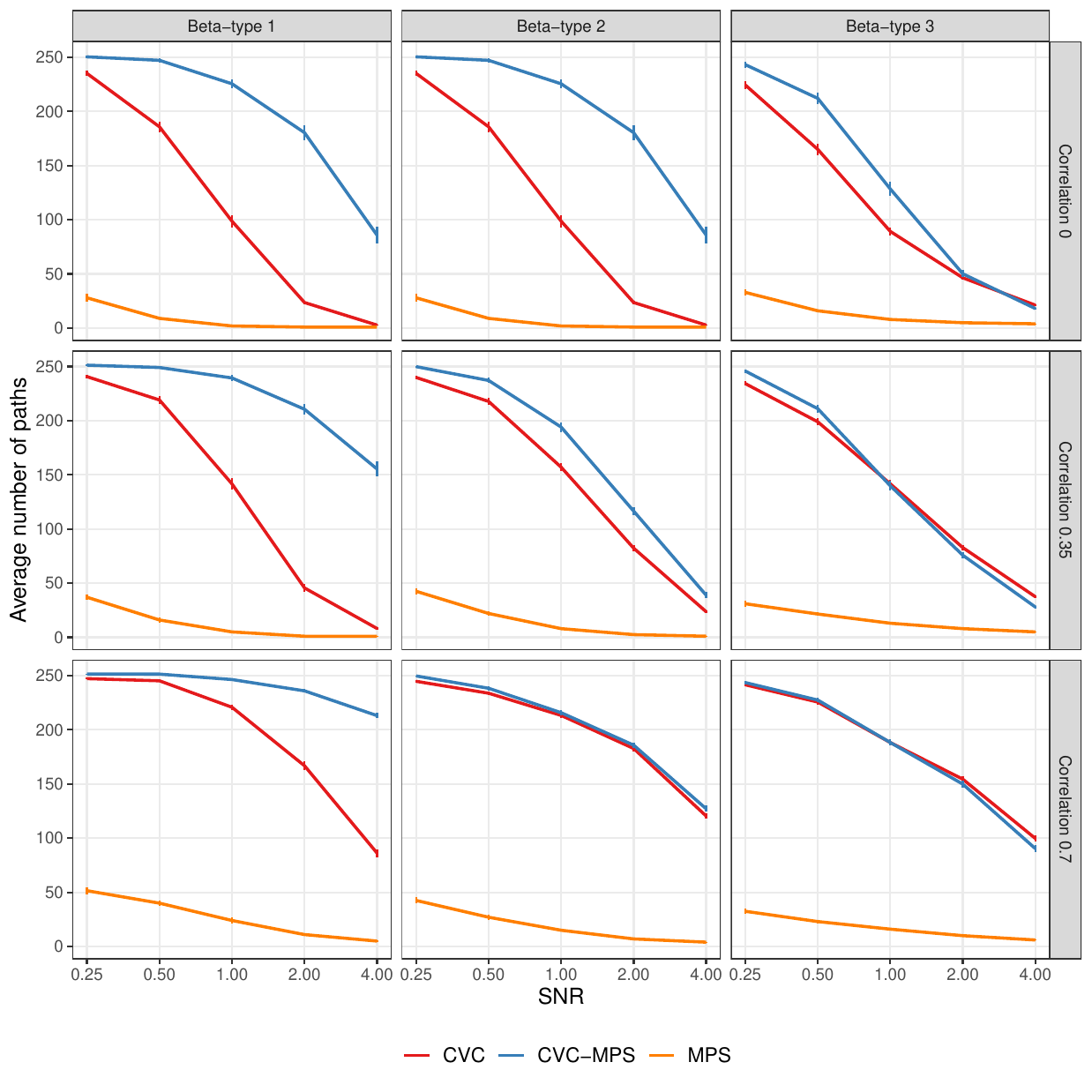}
\caption{Number of models selected by each model set selection method.
\label{fig:sim10size}}
\end{figure}

Figure \ref{fig:sim10} shows the RTEs of all methods for Setup 1 across the beta-types, autocorrelation levels, and SNRs.  Generally, the MSS methods (MPS, CVC, and CVC-MPS) outperform all standard single model selection methods except for some settings at the largest SNRs where the performances converge. The dominance of MSS methods is not particularly surprising given that (i) the MSS methods may select multiple models and (ii) we display only the average minimal observed RTEs from among the selected model sets.  This result does, however, suggest that in most cases, there are alternative models that perform at least as well as those identified by the (single) model selection procedures. These results \emph{do not} imply that \emph{all} models selected by MSS are better, but rather that much of the time there exists at least one model with superior performance.  (Additional plots provided in the appendix show the percentage of time this occurs; in most cases, with the exception of the oracle model, the MSS procedures are able to find more optimal models at least approximately 80\% of the time.)  This phenomena is echoed by the superior performance of MSS methods over even the oracle regression in many settings, supporting the idea that there may be a collection models ``close" to the data generating model. Indeed, these models are so close that by chance on a given sample, they may outperform a regression on just the signal variables.  The results from Setups 2 and 3 are very similar; these figures are given in the Appendix.  In these setups, the MSS methods still generally outperform all of the single model selection methods, though the oracle is often as good or better than the MSS methods.

The strong performance of the MSS methods however, practically speaking, remains only part of the story.  The fact that these procedures are able to routinely identify models that are more accurate than standard model selection methods is encouraging, but this in and of itself is of little real-world value if they require selecting a very large number of models in order to do so.  Figure \ref{fig:sim10size} shows the average number of selected models for each MSS method for each setting in Figure \ref{fig:sim10}.

The results here are fairly striking.  Note from Figure \ref{fig:sim10} that while there is generally a clear differentiation between the model selection and MSS approaches, among the MSS methods, CVC and CVC-MPS tend to perform slightly better than MPS in terms of finding models that, on average, have lower minimum RTEs.  Figure \ref{fig:sim10size} helps explain why:  in every low SNR setting, both CVC and CVC-MPS are selecting nearly every one of the $\binom{10}{5}=252$ models available!  In contrast, the MPS approach proposed in earlier sections selects only between 25 and 50 in those same settings -- a more than 80\% decrease.  Even at high SNRs, both CVC and CVC-MPS often select at least double the number of models selected by MPS.  MPS thus seems to represent a good trade-off between model performance and the number of models selected.

\begin{figure}[t!]
\includegraphics[width=1\columnwidth]{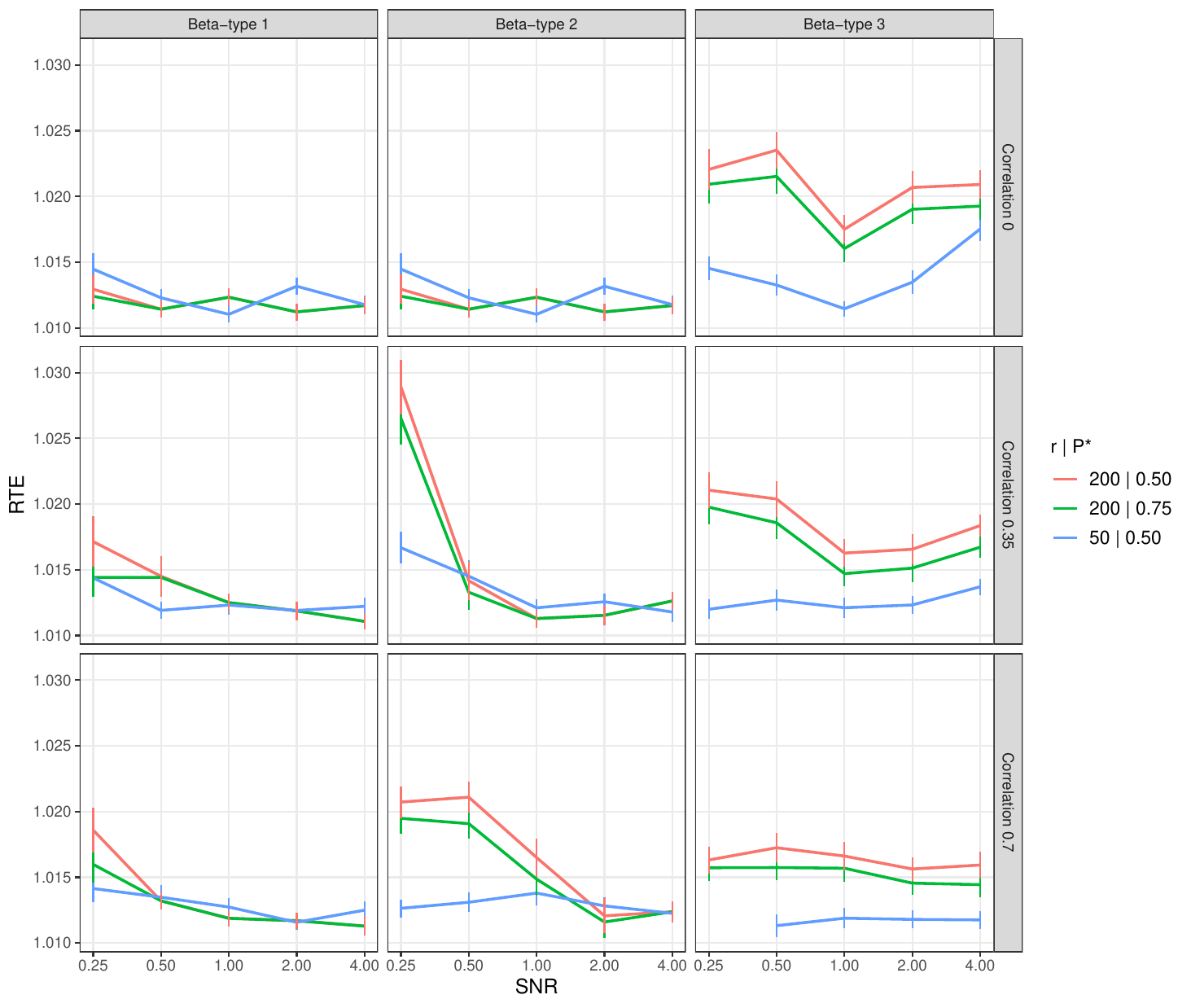}
\caption{RTE of MPS with differing multinomial cell selection rules. Note: the Beta-type 3 and $\rho = 0.7$ plot is a missing datapoint, as its computation time exceeded 6 days. \label{fig:mpscompare_rte}}
\end{figure}

\begin{figure}[t!]
\includegraphics[width=1\columnwidth]{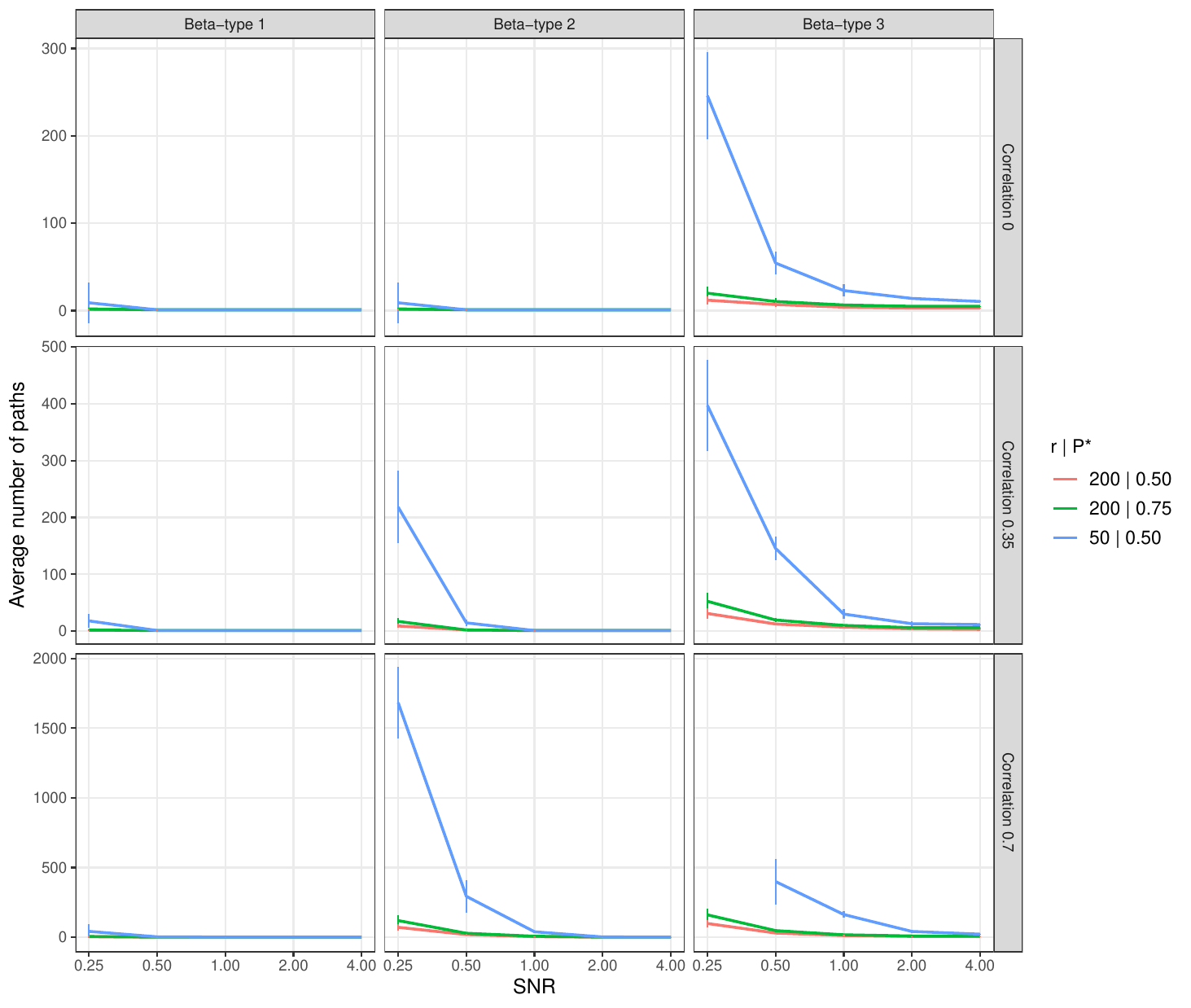}
\caption{Average number of paths of MPS with differing multinomial cell selection rules. Note: the Beta-type 3 and $\rho = 0.7$ plot is a missing datapoint, as its computation time exceeded 6 days.
\label{fig:mpscompare_size}}
\end{figure}

Note also that despite nearly identical performance in terms of minimum RTE, in the beta-type 1 (Figure \ref{fig:sim10size} first column) and 0-correlation (Figure \ref{fig:sim10size} top row) settings, CVC-MPS often selects substantially more models than CVC.  While this is an obvious drawback, we stress that this reduction in model set size comes at the cost of increased computational effort.  Figure \ref{fig:sim10time} in the appendix shows the relative runtimes of these two methods. Here we see that CVC-MPS takes between $\frac{1}{2}$ and $\frac{1}{25}$ the time of CVC, which is particularly impressive considering that in some settings both methods perform nearly identically in terms of both model set size and minimum RTE. The computational gains of CVC-MPS can thus be immense, though these gains can sometimes come at the cost of an increased number of selected models. It should also be noted that in higher-dimensional settings where the number of candidate models is large, CVC may be entirely computationally prohibitive.

To close out our simulations, we turn our attention exclusively to the MPS procedure defined in Algorithm \ref{alg:mps2} and examine the impact of different selection rules.  Here we fix $n=500$, $p=100$, and $s=5$ as in Setups 2 and 3 and perform MPS with parameter pairs ($r = 200, P^* = 0.75$), ($r = 200, P^* = 0.50$), and ($r = 50, P^* = 0.50$).  

Average minimum RTEs and selected model set sizes are shown in Figures \ref{fig:mpscompare_rte} and \ref{fig:mpscompare_size}, respectively.  Note that in these figures, the ($r = 50, P^* = 0.50$) datapoints are averaged over different datasets from ($r = 200, P^* = 0.75$) and ($r = 200, P^* = 0.50$).  We see that fixing $r = 200$ and lowering $P^*$ from $0.75$ to $0.50$ results in a slightly inflated minimum RTE and a correspondingly small decrease in the average number of models selected.  This should come as no surprise:  these probabilities are akin to confidence levels and thus decreasing $P^*$ means that fewer covariates are selected at each step, leading to fewer overall models being selected.  On the other hand, keeping $P^* = 0.50$ and lowering $r$ from $200$ to $50$ reduced the minimum RTE in several cases and greatly increased the average number of models selected -- the selection rule for $r = 50$ is more conservative than when $r = 200$. In effect, the value of $r$ operates in an analogous fashion to the the sample size in a confidence interval calculation, tightening the width of the interval at larger values. By lowering the value of $r$, we obtain considerably more paths -- in some cases by more than tenfold -- and despite also lowering $P^*$, maintained similar if not better RTE results. Of course, the consequence of these better RTE results come at the cost of increased paths, which can exceed $1{,}500$, though it is worth noting that this is still less than 0.002\% of the $\binom{100}{5} = 75{,}287{,}520$ candidate models available.

\section{Real-World Data Applications}
\label{sec:applications}
We now demonstrate the value of the MPS procedure on two popular real-world datasets.  The first example illustrates that even with relatively simple data, MPS can sometimes identify a large number of similarly-accurate models, many of which outperform the individual models selected via forward selection and the lasso.  The second example demonstrates that the number and diversity of optimal models selected depends heavily on the model class and loss function specified.

\subsection{Diabetes Disease Progression}

We first examine the diabetes dataset provided in \citet{Efron2004}, which was originally provided to demonstrate least-angle regression (LARS) in a sparse linear regression regime but more recently has been used to demonstrate the behavior of CVC \citep{Lei2020}. The data describe $ n = 442 $ diabetes patients who have ten baseline measurements recorded: age, sex, body mass index (bmi), mean arterial pressure (map), and six blood serum measurements (tc, ldl, hdl, tch, ltg, glu), all of which are standardized.  In addition, a response variable measuring diabetes progression one year after baselines were recorded is also given. As was done in both \citet{Lei2020} and \citet{Efron2004}, we consider all second order terms, thus bringing the total number of predictors to $p = 64$. We also randomly split the data into a training and test sets of sizes $n=300$ and $n=142$, respectively.

\begin{figure}[t!]
\includegraphics[width=0.86\columnwidth]{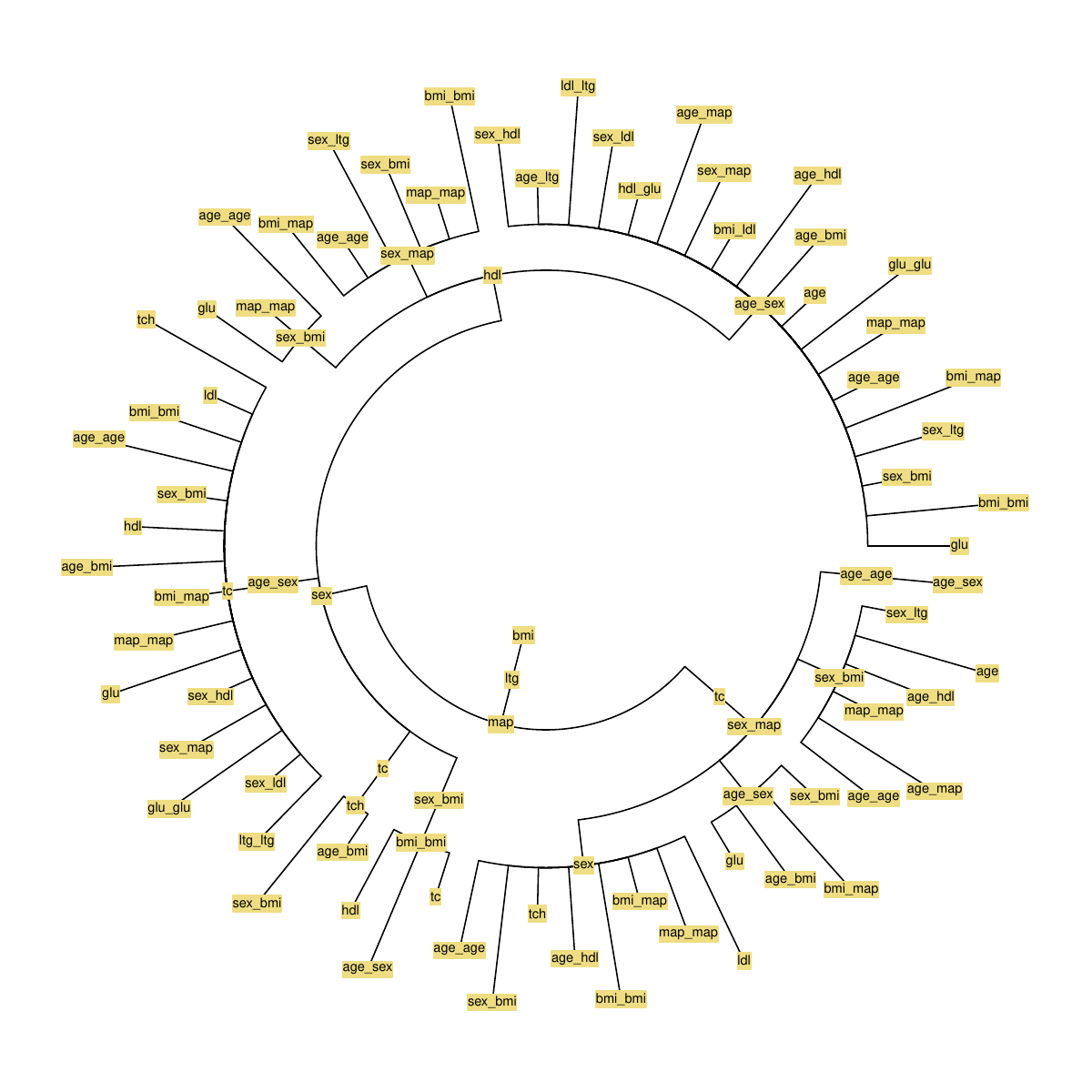}
\caption{Model paths of diabetes dataset plotted radially; there are a total of 67 paths. At a depth of 7, the number of paths explodes, suggesting that forward selection becomes relatively unstable at models of that size.
\label{fig:diapaths}}
\end{figure}

\begin{figure}[]
\includegraphics[width=0.7\columnwidth]{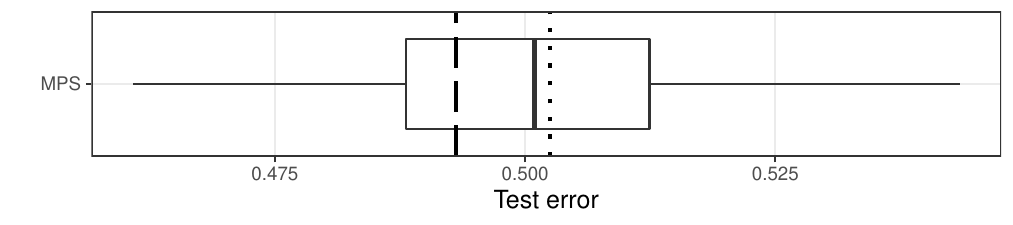}
\caption{Box plot displaying test errors of models selected via MPS.  The vertical dashed line to the left shows the test error of forward selection; the vertical dotted line to the right shows the test error of lasso.
\label{fig:diaerr}}
\end{figure}

In keeping with what is common in applied research areas, we begin by performing both lasso and forward selection, both of which are tuned via 5-fold cross validation. MPS is also performed using linear models with squared-error loss where we set an inclusion probability of $P^* = 0.95$, a maximum cell count of $r = 100$, and a depth matching that of the cross-validated forward selection. The selected model paths are shown in Figure \ref{fig:diapaths}.  In analyzing the selected paths, we see that a total of 67 models were selected, thus demonstrating that there indeed are numerous well-performing models -- many of which outperform the individual models selected via both lasso and forward selection. In fact, of the 67 models identified 24 (36\%) are more accurate on our test set than the lasso model and 40 (60\%) are more accurate than the linear model obtained via classic forward selection.  Figure \ref{fig:diaerr} shows a box plot of the model errors from MPS with the lasso and forward selection models overlaid for reference.  At the very least, this suggests that models selected via forward selection or lasso should not be treated as the only well-performing models, and any discussion about the particular importance of covariates selected in such models would be misguided.  This stands in contrast to most applied research articles, wherein authors perform model/variable selection and then carry out inferential procedures as though the resultant model was chosen \textit{a priori}. 

\subsection{Breast Cancer Identification}

We now consider a dataset describing cell nuclei images taken from 699 breast cancer patients \citep{Bennett1992}. The variables present are the clump thickness (X1), uniformity of cell size (X2), uniformity of cell shape (X3), marginal adhesion (X4), single epithelial cell size (X5), bare nuclei (X6), bland chromatic (X7), normal nucleoli (X8), mitoses (X9). The binary response indicates whether or not the measured tumor was determined to be malignant.

The purpose of this application is to show the behavior of MPS under different modeling conditions and demonstrate how model class heavily impacts the models identified and the covariates utilized. We apply MPS once with regression trees and again with logistic regression, both with squared error loss.  Given the complexity of the underlying system, it is conceivable that logistic regression is too coarse a model to uncover unique meaningful relationships and it was thus hypothesized that model selection procedures may be more unstable relative to regression trees. MPS is performed with an inclusion probability of $P^* = 0.75$, maximum cell count of $r = 200$, and a depth of 3. 

\begin{figure}[]\centering
\includegraphics[width=1\columnwidth]{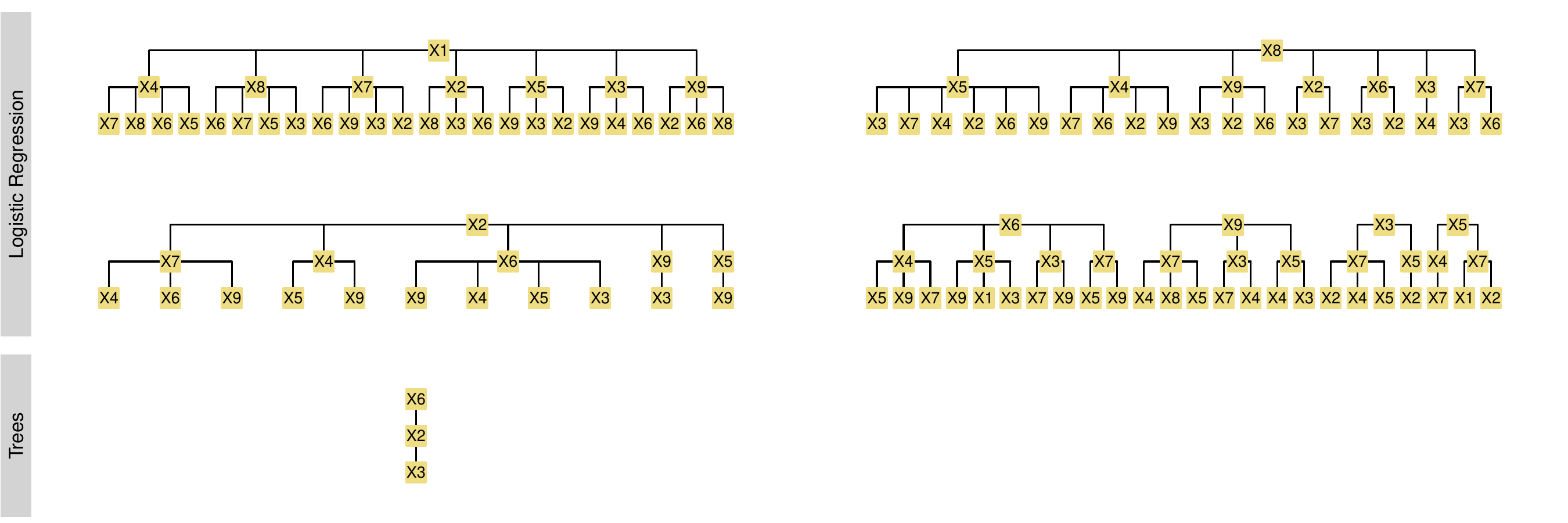}
\caption{Path structure for logistic regression (top 2 rows) and regression trees (bottom). MPS produces 79 paths using logistic regression and only 1 path using regression trees.
\label{fig:breastpaths}}
\end{figure}

The resultant model paths are shown in Figure \ref{fig:breastpaths}.  Disappointingly, we do indeed observe that logistic regression produces a very large number of plausible models.  In fact, logistic regression produces 79 models -- more than 94\% of the $\binom{9}{3}$ = 84 unique 3-variable models available.  Shockingly, MPS with regression trees produces only 1 single path, suggesting that this particular modeling context is very stable.

\section{Discussion}
\label{sec:discussion}

The model path selection (MPS) framework introduced here provides scientists and practitioners with a computationally-efficient tool allowing them to graphically visualize the stability of the modeling process.  The procedure itself is greedy in the sense that models are built up via forward selection, but that greed is mitigated by the fact that all plausibly-optimal variables are included as possibilities at each step.  We stress as well that MPS should be seen as a generic wrapper-style method and in particular, we make no claims of optimality with respect to the particular implementation utilized here in which decisions at each step are made via classical results from the ranking-and-selection literature.  Future work may explore more computationally advantageous approaches, especially in high-dimensional settings wherein it may be reasonable to implement a kind of ``screening"  or ``early stopping" rule that would quickly eliminate covariates at each step that are unlikely to be chosen in the final set.

In contrast with many other procedures, MPS is completely agnostic to the model and loss function employed and users may choose to construct models to whatever depth is desired.  Thus, in practice, the model depth could be constrained for practical purposes or chosen in a data-dependent fashion, such as by some form of validation or by stopping once the empirical gain in model accuracy that could be obtained by adding more variables drops below some threshold. This flexibility is also potentially of benefit from a theoretical perspective.  It is reasonable, for example, to ask how likely it is that the ``true" model will be found via MPS if it is assumed to exist and be discoverable via forward selection.  Such questions, however, depend entirely on the particular model, loss, and stopping rules employed. 

It may be apparent to some readers that our MPS procedure described above is merely one possible form of a forward selection analogue of MSS.  As an alternative, one could, for example, perform MSS on \emph{all} $j$-variable models that nest \emph{all} models in $\mathcal{M}^*_{j-1}$, rather than performing a separate MSS for the $j$-variable models that nest each $m \in \mathcal{M}^*_{j-1}$. Certainly, this is a valid possible alternative, though we stress that this kind of approach may still result in a large number of model comparisons and thus be computationally overbearing. Under our formulation of MPS, the intermediary candidate sets are broken up into numerous smaller sets so to prevent burdensome computational issues. 

Finally, we note that there are a number of minor extensions of the basic MPS procedure that could readily be employed in practice.  If one is interested in groups rather than individual covariates, entire groups may be identified at each step.  Similarly, depending on the particular modeling framework being used, it may be advantageous to employ an adaptive form of MPS wherein interactions would be allowed in later steps once individual covariates are selected.  Lastly, if one is particularly concerned about ensuring that none of the models selected via MPS are significantly less accurate than the others, an exhaustive search procedure like CVC could be run on that final collection of models.  Indeed, especially when many models are produced, it is reasonable to worry about a kind of propagation of error -- perhaps at each step there isn't enough evidence to exclude a particular variable, but there may be evidence at the end of the process that a model consisting of many of the least-frequently chosen variables at each step is ultimately suboptimal.  If MPS is seen as constructing model ``trees" of sorts, then applying a CVC-type filter after construction might be seen as something akin to pruning.  

\section*{Acknowledgements}
This research was supported in part by the University of Pittsburgh Center for Research Computing through the resources provided.  LM was supported by NSF DMS-2015400.  We would also like to thank Will Fithian for a helpful early conversation regarding the ranking and selection literature. 

\bibliographystyle{apalike}
\bibliography{database}

\newpage

\appendix
\section{Additional Simulation Results}
\begin{figure}[]
\includegraphics[width=0.84\columnwidth]{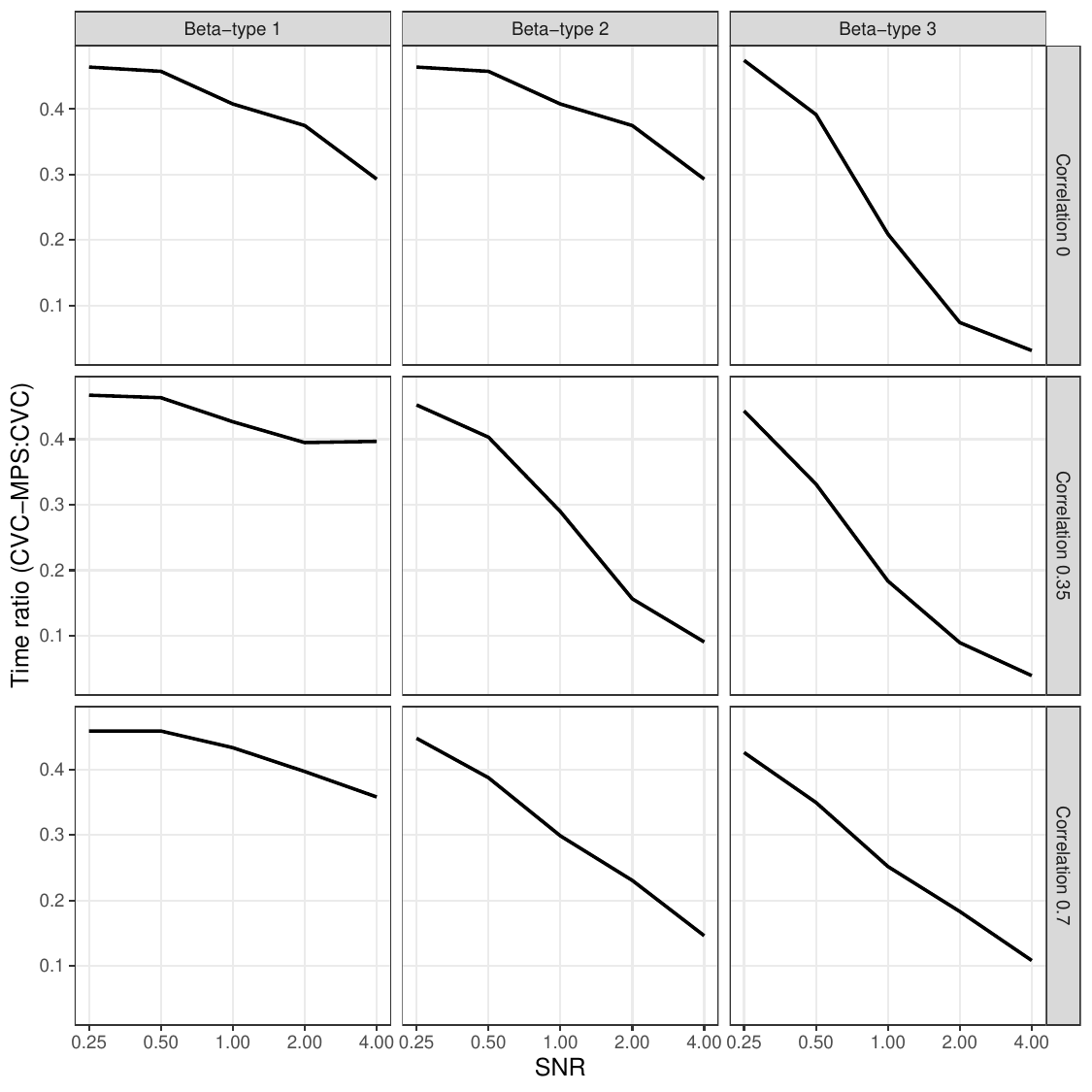}
\caption{Ratio of average computation time of CVC-MPS to CVC for Simulation Setup 1. 
\label{fig:sim10time}}
\end{figure}

\begin{figure}[]
\includegraphics[width=0.84\columnwidth]{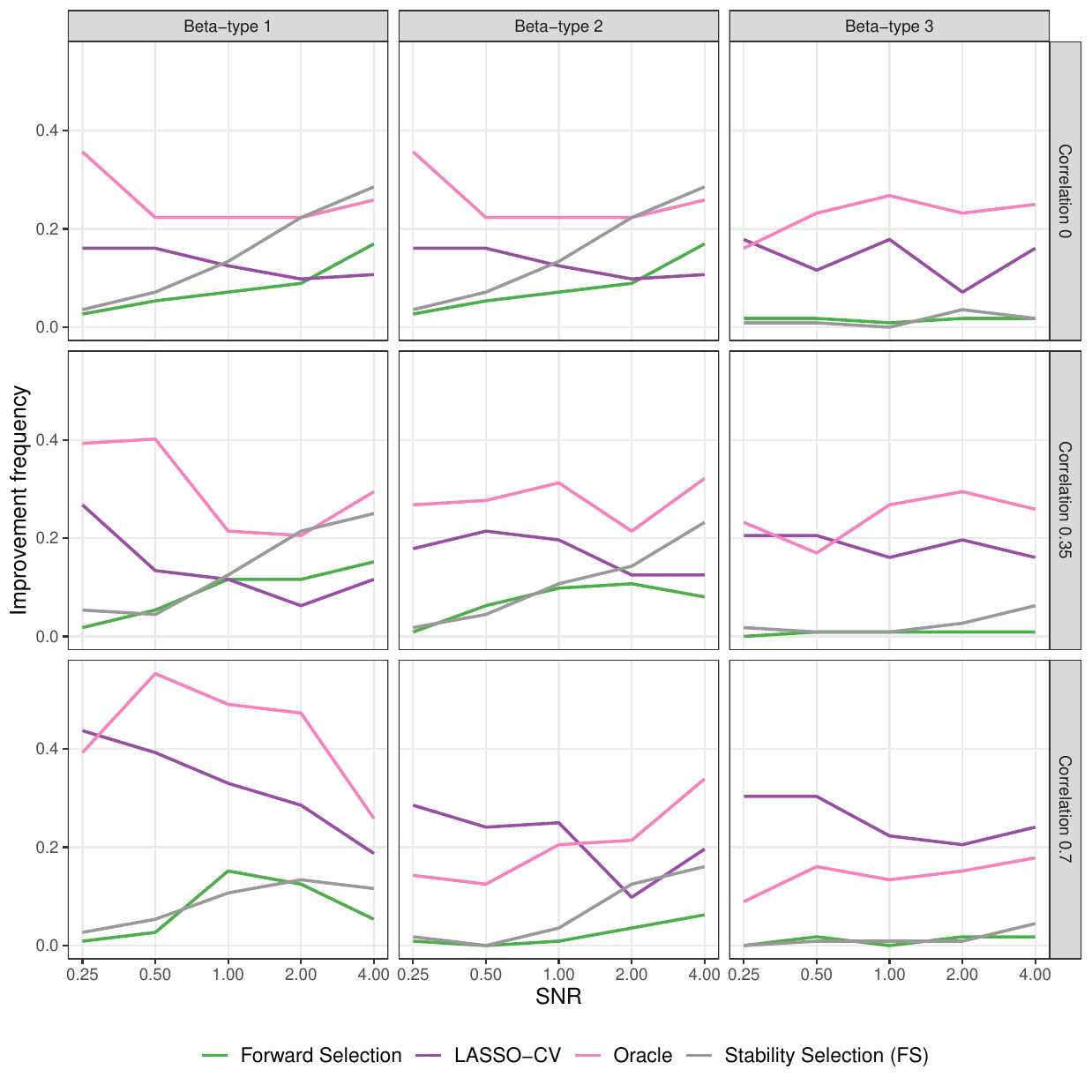}
\caption{Proportion of times each model selection method obtains a lower test RTE than all models selected via MPS for Setup 1 ($p=10$, $r = 200$, $P^* = 0.95$). 
\label{fig:propwin10_r200}}
\end{figure}

\begin{figure}[]
\includegraphics[width=0.84\columnwidth]{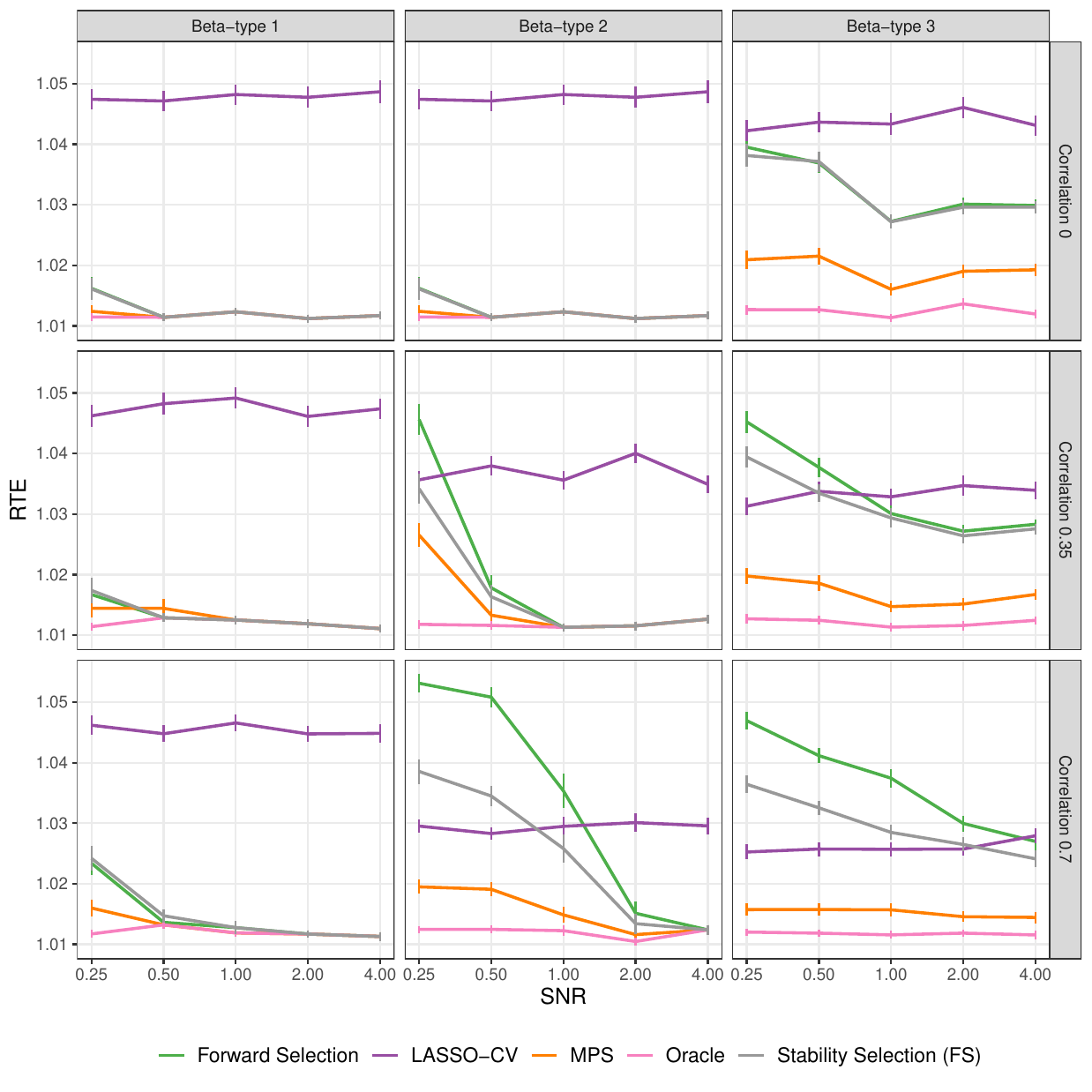}
\caption{RTE of each model selection type for Setup 2 ($p=100$, $r = 200$, $P^* = 0.75$). 
\label{fig:sim100_r200}}
\end{figure}

\begin{figure}[]
\includegraphics[width=0.84\columnwidth]{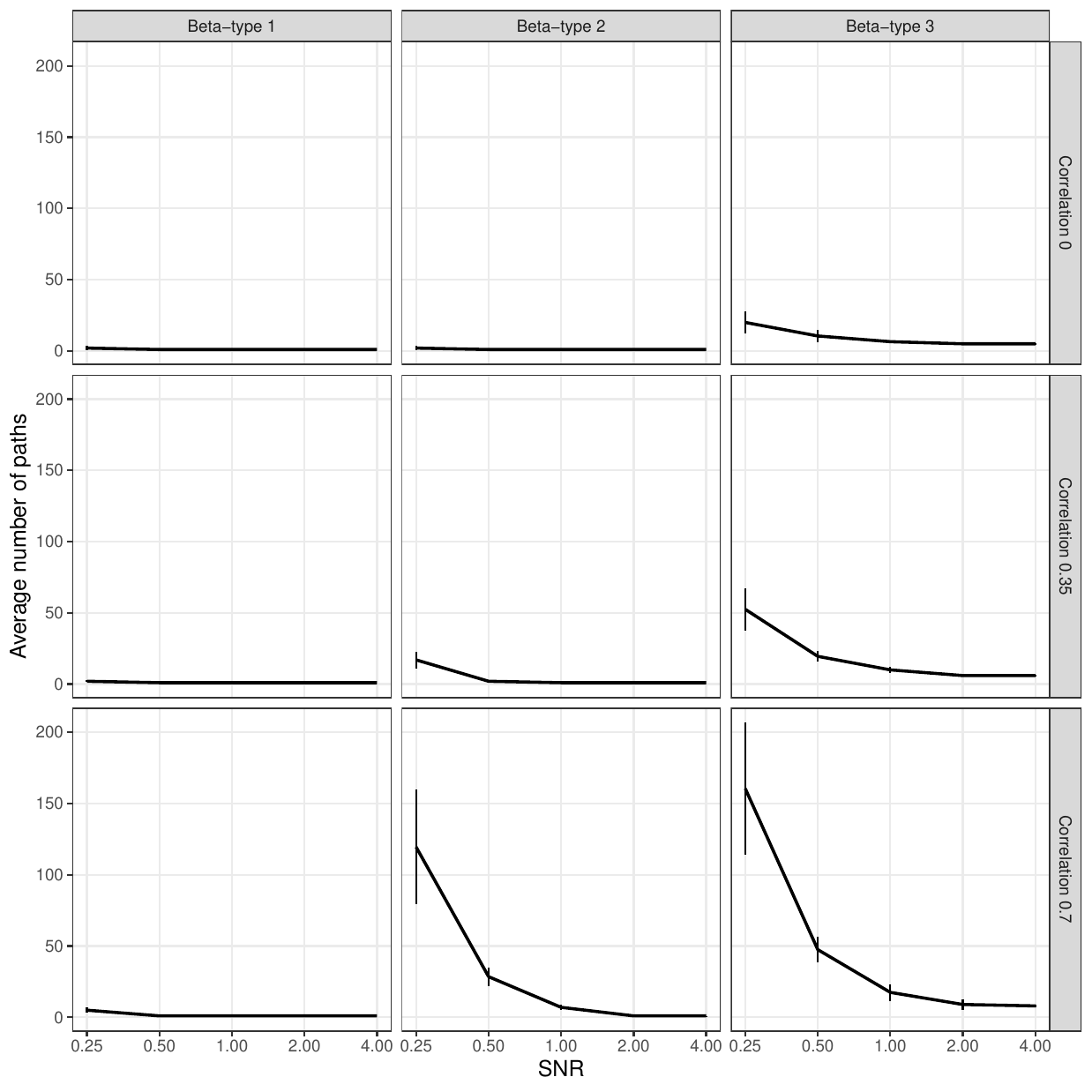}
\caption{Number of paths for MPS in Setup 2 with $p=100$, $r = 200$, $P^* = 0.75$. 
\label{fig:sim100_r200size}}
\end{figure}

\begin{figure}[]
\includegraphics[width=0.84\columnwidth]{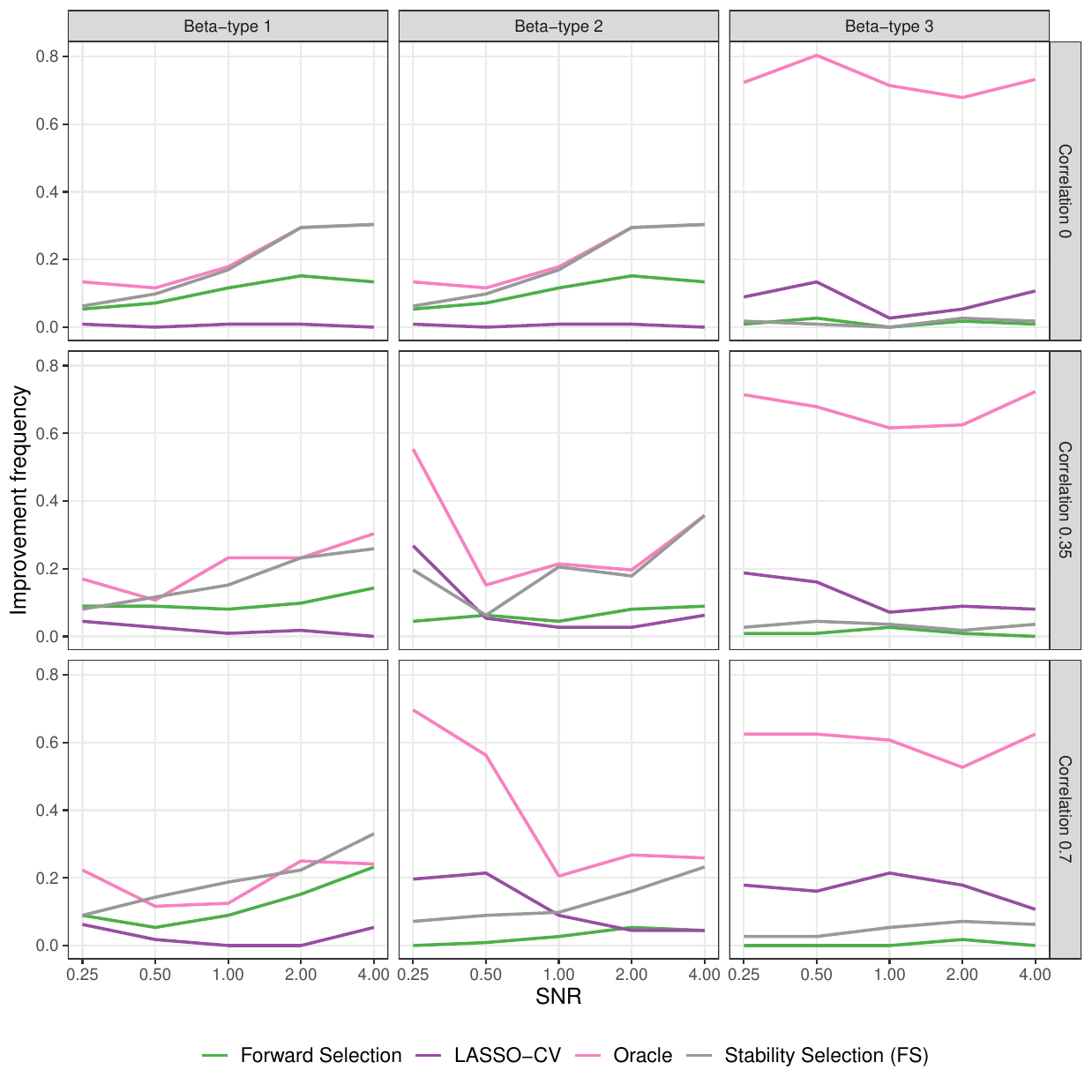}
\caption{Proportion of times each model selection method obtains a lower test RTE than all models selected via MPS for Setup 2 ($p=100$, $r = 200$, $P^* = 0.75$). 
\label{fig:propwin100_r200}}
\end{figure}

\begin{figure}[]
\includegraphics[width=0.84\columnwidth]{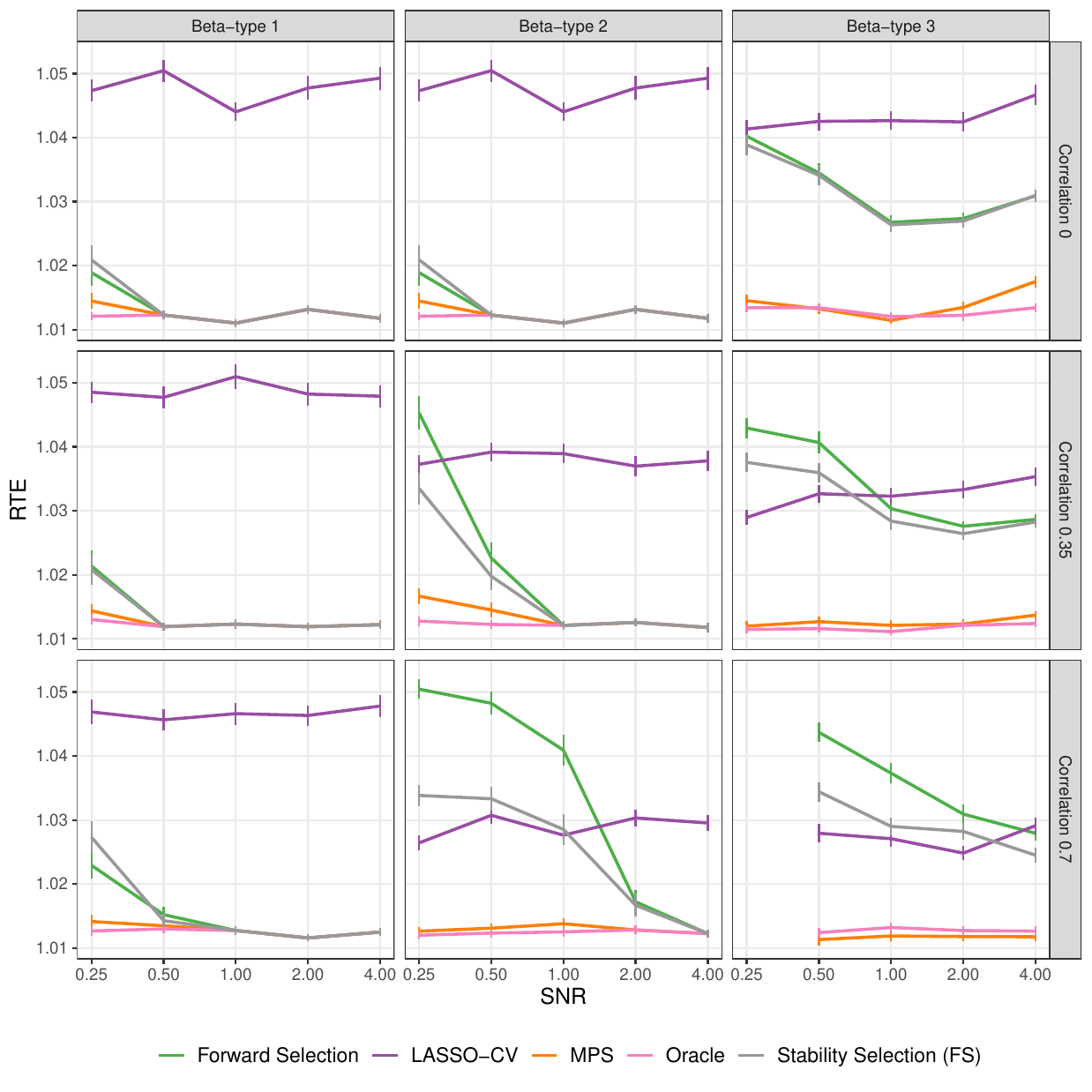}
\caption{RTE of each model selection type for Setup 3 ($p=100$, $r = 50$, $P^* = 0.50$).
\label{fig:sim100_r50}}
\end{figure}

\begin{figure}[]
\includegraphics[width=0.84\columnwidth]{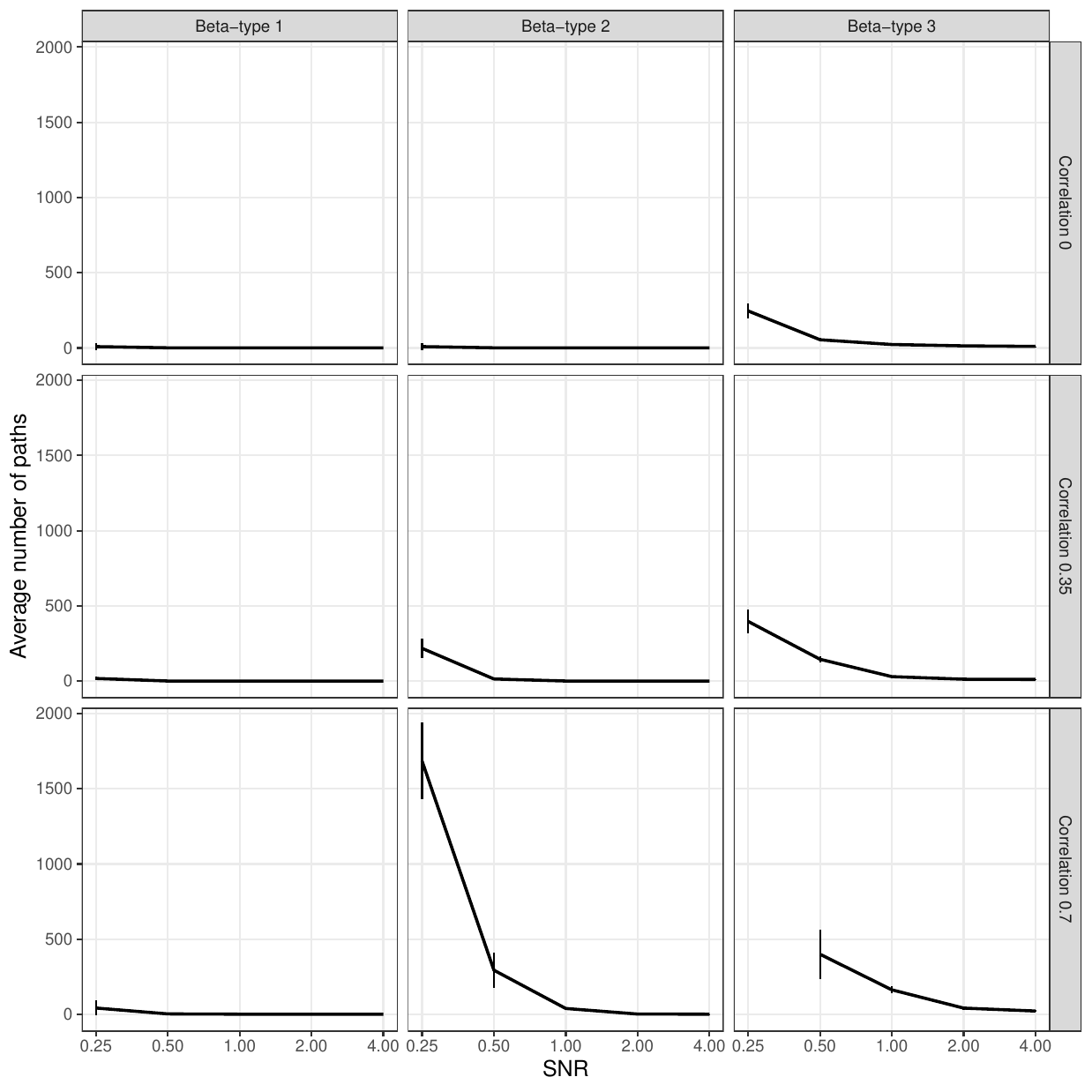}
\caption{Number of paths for MPS in Setup 3 with $p=100$, $r = 50$, $P^* = 0.50$. 
\label{fig:sim100_r50size}}
\end{figure}

\begin{figure}[]
\includegraphics[width=0.84\columnwidth]{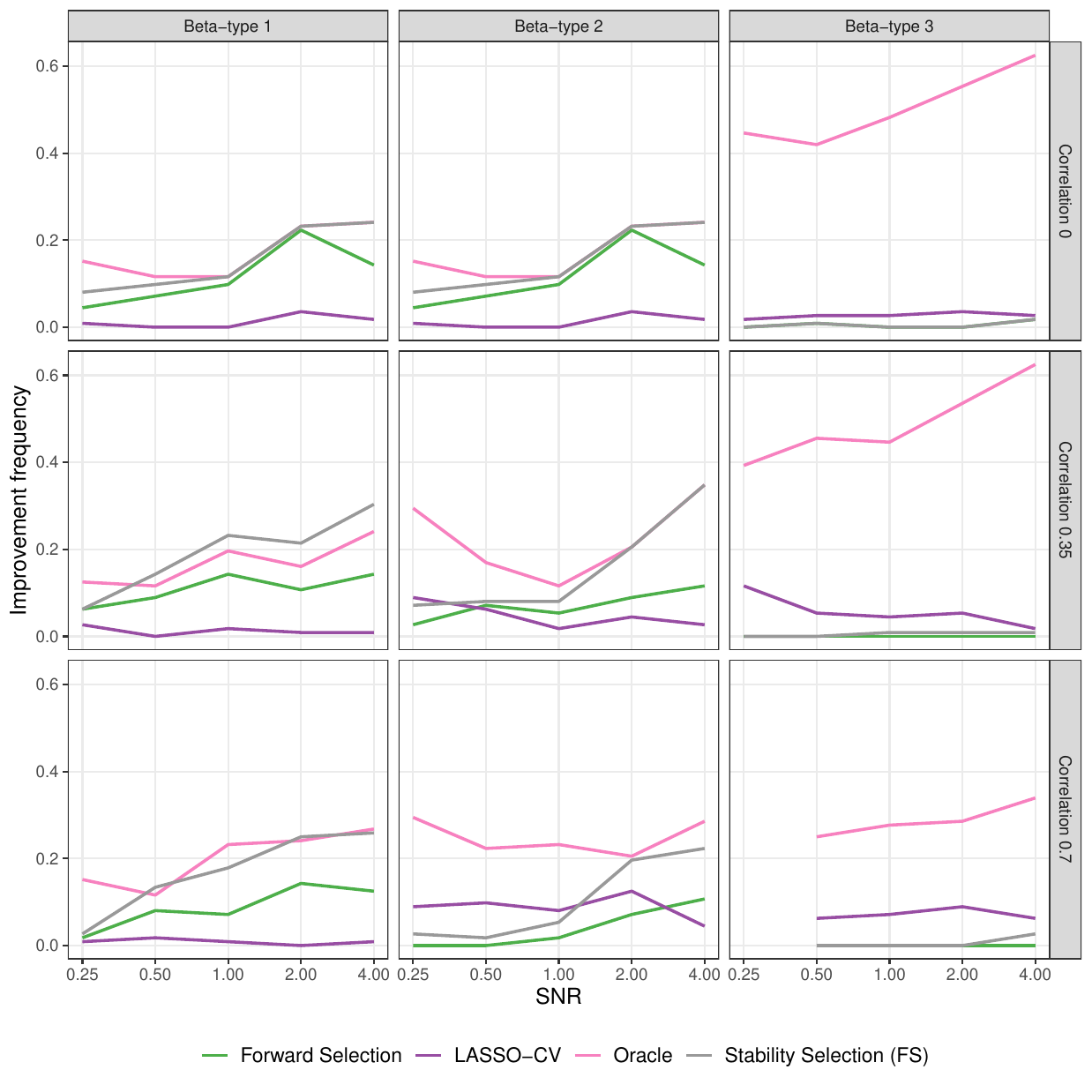}
\caption{Proportion of times each model selection method obtains a lower test RTE than all models selected via MPS for Setup r ($p=100$, $r = 50$, $P^* = 0.50$). 
\label{fig:propwin100_r50}}
\end{figure}

Here we present a number of additional figures corresponding to simulation results from Section \ref{sec:simulations}.  Figure \ref{fig:sim10time} shows the ratio of computational times for CVC-MPS to CVC in Simulation Setup 1.  The results -- average minimum RTE and average number of models selected -- for Setup 2 are given in Figures \ref{fig:sim100_r200} and \ref{fig:sim100_r200size}, respectively; results for Setup 3 are given in Figures \ref{fig:sim100_r50} and \ref{fig:sim100_r50size}, respectively.  Note that in these latter two setups, we do not perform CVC or CVC-MPS due to the increased computational burden. Additionally, we provide plots that display the frequency with which each model selection method outperforms all models selected via MPS in Figure \ref{fig:propwin10_r200}, Figure \ref{fig:propwin100_r200}, and Figure \ref{fig:propwin100_r50}. \\

\end{document}